# Investigation on Vacuum Field Theory

## G. Gemedjiev

# KINEMATICS OF THE VACUUM FIELD THEORY

G. Gemedjiev

Plovdiv University "Paisii Hilendarski", 24 Tsar Asen Str. 4000 Plovdiv, Bulgaria
e-mail: gemedjievg@mail.bg

*The vacuum field theory is the conventional denomination of a new field theory based on two axioms concerning two basic states of vacuum. According to the first one, which will be applied in kinematics, dynamics and electrodynamics, when neither gravitational, electromagnetic nor other fields are acting, vacuum in a confined domain of space round the connected to the same vacuum inertial reference system and in the scope of a given inertial reference system, is in an inactivated state. This state is characterized in both cases with isotropic propagation of light at a constant speed $c$. On the basis of the first axiom and the presumption that the physical phenomena in the macrouniverse are space and time defined, using the conventional definitions of the measurement operations in the inertial reference systems, moving relative to each other, the simultaneity of events in these frames has been proved. A model based on the experiments is proposed to present the interaction between vacuum and the material bodies. It entails that they do not change their dimensions due to their inertial motion in relation to vacuum or other material bodies, i. e. it is concluded that the space is a Newtonean space. The transformation of the coordinates, the relative velocities and the time during the transition from the vacuum connected reference system $K$ to the inertial reference system $K'$ moving relative to the first one and the opposite transition are determined on the basis of the above model and the axiom. Based on the axiom concerning the inactivated state of the vacuum, the same transformation are valid for the motion of the inertial reference system $K''$ in the proper space of frame $K'$ on condition that its mass and dimensions are much smaller than those of frame $K'$. The transformation, mentioned above, are used to formulate the law of velocity summation and to derive the formulas of the four-dimensional vector, velocity and acceleration. The law of velocity summation was used to investigate the propagation of light in an optically denser moving environments and, as a result, the formula of Fizeau was derived. The famous Thomas formula of precession has been obtained. It has been shown that the proper time in the inertial reference system $K'$ is an invariant.*

## 1. Simultaneity in inertial reference systems moving relative to each other

Let us assume that at a moment of time the axes of the inertial reference system $K'$ coincide with the respective axes of the inertial reference system $K$. We shall assume that frame $K'$ is moving relative to frame $K$ with a constant velocity $\vec{v}$. The direction of the velocity coincides with the direction of axis $OX$ in frame $K$. A unit immovably connected to frame $K$ will be used to prove the simultaneity of the events in frames $K$ and in $K'$. Its cross-section in the plane $O'X'Y'$ is presented in Fig. 1. The unit consists of a cylindrical tube closed at both ends. Two glass cylinders are installed symmetrically towards the centre of the tube $Q'$ and towards its two ends. The outer matted bases of the cylinders are perpendicular to their axes while the inner ones make an angle of $45^\circ$ with them. They are oriented in such a way, that after being scattered by the matted bases, the part of the light signal propagating along their axis turns off and propagates along the axis of the tube towards its center $Q'$. The unit is vacuumed. The speed of light in vacuum relative to frame $K'$ is $c$ and along the axes of the glass cylinders it is $c'$. Let us assume that plane $O'Y'Z'$ which crosses the center of the tube $Q'$ perpendicularly to its axis is a plane of symmetry for the unit.

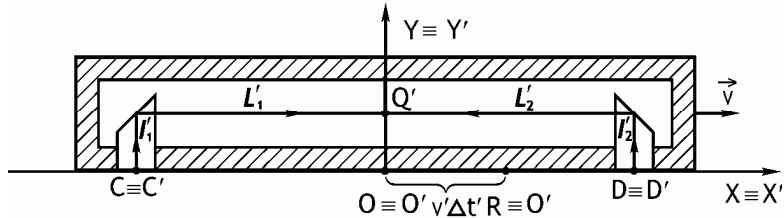

Fig. 1.

Let at a moment of time $t_o$, read in frame $K$, its points $C$ and $D$, and the point $O$, situated in the middle between them, coincide with points $C'$, $D'$ and the point $O'$ situated in the middle of frame $K'$. The time is measured by synchronized watches situated near points $C$ and $D$ from which at time $t_o$, light signals are sent in the direction of the tube axis. Since the plane $O'Y'Z'$ is a plane of symmetry for the unit, the signals from points $C$ and $D$ will reach point $Q'$ in the same interval of time $\Delta t'$ equal to

$$\Delta t' = \frac{l'_1}{c'} + \frac{L'_1}{c}. \tag{1}$$

Point $O'$ in frame $K'$ passes the distance $r'$, for the same interval of time, which is measured in frame $K'$ and is equal to

$$r' = v' \Delta t' . \qquad (2)$$

where $v'$ is the speed of frame $K$ relative to frame $K'$. At the end of the time interval $\Delta t'$ the position of point $O'$ will coincide with that of point $R$ of frame $K$.

As is known from the measurement operations a period of time $\Delta t$ measured in frame $K$ while point $O'$ is passing by points $O$ and $R$ in frame $K$, corresponds to a certain period of time $\Delta t'$, measured in frame $K'$. Consequently, relative to frame $K$, the two signals will reach point $Q'$ at the same moment when its clocks will read

$$t = t_o + \Delta t . \qquad (3)$$

In the vacuum field theory we assume, that in the macrouniverse the axiom for the space and time definition of the position of a certain point $S$ is in force. This axiom, applied to any two inertial reference systems $K$ and $K'$, moving relative to each other, states: if at a certain moment of time $t$ the position of point $S$ in frame $K$ can be presented as $S(x, y, z, t)$, the position of point $S$ in frame $K'$ at the same moment will be $S(x', y', z', t')$.

From the axiom concerning the space and time definition, it follows, that to the moment of time $t$ in frame $K$ for point $Q'$ in frame $K'$ corresponds only one moment of time $t'$. Consequently, relative to frame $K'$, both signals will reach point $Q'$ at a moment of time $t'$. This is possible only in case when the two watches in points $C'$ and $D'$ in frame $K'$ read at the beginning of the motion one and the same time $t'_o$, which is equal to:

$$t'_o = t' - \Delta t' , \qquad (4)$$

i.e. the simultaneous events in frame $K$ are simultaneous in frame $K'$ as well.

Let by a measurement operation in frame $K$ be determined that to the length $\ell'$ in frame $K'$ there corresponds the length $\ell$ in frame $K$. It follows directly from the simultaneity of the events in frames $K'$ and $K$ that the lengths $\ell'$ and $\ell$ are measured at the same moment of time in both the frame $K'$ and frame $K$. Consequently, the ratio $\dfrac{\ell'}{\ell}$ will be one and the same no matter in which of the frames $K$ or $K'$ it is measured. Therefore, with proven simultaneity of the events in the inertial reference systems in the macrouniverse, we shall deal with measuring the real deformations if they actually exist. The indirect methods of measuring the refraction index of a moving body, used in the experiments of Rayleigh [1] and Brace [2] did not establish the existence of real deformations. The experiments of Trouton and Rankine [3], measuring the electrical resistance of a moving conductor, and those of Wood, Tomlison, and Essex [4], measuring the frequency of an oscillating quartz rod, have come to the same conclusion.

## 2. Vacuum model in an inertial reference system

With the vacuum model which will be used we shall assume that when the bodie move in vacuum, it influences them. Its influence is equivalent to that of material particles $m_v$ much smaller than the atom. They will not influence an inertial reference system only if their average velocity relat to the frame equals zero. This frame will be called a proper reference system of the vacuum in the space domain discussed and will be denoted by frame $K$.

For the above-mentioned model the density of $m_v$ particles is presumed to be the same in the space volume of frame $K'$ and their average velocity relative to the same frame to be equal to zero. As a result of the motion of frame $K'$ relative to frame $K$ the density of $m_v$ particles in the first frame will be considered higher than their density in the second one. These assumptions make it obvious, that the propagation of light in the proper space of frame $K'$ will propagate isotropically and with a constant speed. Another consequence for the above-mentioned model is that if some deformations appear as a result of the bigger density of $m_v$ particles in frame $K'$ as a macrobody, they will be identical in all the directions because of the same density of the $m_v$ particles in the proper space of frame $K'$ in its volume. From now on, this volume will be denoted by a proper space of frame $K'$.

Due to the greater density of $m_v$ particles of frame $K'$ physical processes in it will be delayed and time in it will go slower than in frame $K$. This enables us to determine the speed of frame $K'$ relative to the inactivated state of vacuum in a restricted spatial domain. This is carried out by defining in which proper inertial frame $K$ time goes faster compared to any other inertial reference system $K'$, irrespective of the magnitude and the direction of its velocity relative to frame $K$.

## 3. Kinematic characteristics of the model of the interaction of the inertial reference systems with vacuum

Let us assume that the motion of frame $K'$ relative to frame $K$ is accomplished in the standard way, described in Sec. 1 and the coordinates of point $P'$ in frame $K'$ are $x'$, $y'$, $z'$, while in frame $K$ the coordinates of the same point are $x$, $y$, $z$. With this model the transformation of the coordinates of point $P'$ at a moment of time $t$ in frame $K$ corresponding to a moment of time $t'$ in frame $K'$ is expressed by:

$$x = \gamma_x(x' + v't'), \quad y = \gamma_y y', \quad z = \gamma_z z', \tag{5}$$

where $\gamma_x$, $\gamma_y$ and $\gamma_z$ are coefficients presenting the deformation of the coordinate axes of frame $K'$ relative to the respective coordinate axes of frame $K$. No deformations were established in the experiments mentioned in item 1. The coefficients $\gamma_x$, $\gamma_y$ and $\gamma_z$ are identical in the model accepted in Sec. 2. So:

$$\gamma_x = \gamma_y = \gamma_z = 1. \tag{6}$$

If it is accepted that when the origins of the two frames have coincided the clocks will read a zero moment of time then for an interval of time $t$ the origin of frame $K'$ should have moved relative to the origin of frame $K$ at a distance of $\Delta x = vt$. The same distance, measured in frame $K'$ using static scales, will be equal to $\Delta x' = v't'$. The distances $\Delta x$ and $\Delta x'$ are measured simultaneously and

$$\Delta x = \gamma_x \Delta x' = \Delta x' \tag{7}$$

hence

$$t = \frac{v'}{v} t'. \tag{8}$$

Formulas (5) and (8) express the transformation of the coordinates and time in the transition from frame $K'$ to frame $K$. In the same way the transformation of the coordinates and time in the transition from frame $K$ to frame $K'$ can be expressed thus

$$x' = x - vt, \quad y' = y, \quad z' = z, \quad t' = \frac{v}{v'} t. \tag{9}$$

### 4. Law of velocity summation

Differentiating (5) with respect to $t$ and taking into account that

$$\frac{dt'}{dt} = \frac{v}{v'} \tag{10}$$

the components of the velocity $V_x$, $V_y$, $V_z$ of point $P'$ relative to frame $K$, expressed through the components of its velocity $V'_x$, $V'_y$, $V'_z$ relative to frame $K'$, will be

$$V_x = \frac{dx}{dt} = \frac{dx}{dt'} \frac{dt'}{dt} = \frac{v}{v'}(V'_x + v'), \quad V_y = \frac{v}{v'} V'_y, \quad V_z = \frac{v}{v'} V'_z. \tag{11}$$

Analogously from transformation (9) for the components of the velocity $V'_x$, $V'_y$, $V'_z$ of point $P$ relative to frame $K'$, expressed through the components of its velocity $V_x$, $V_y$, $V_z$ relative to frame $K$, we find

$$V'_x = \frac{v'}{v}(V_x - v), \quad V'_y = \frac{v'}{v} V_y, \quad V'_z = \frac{v'}{v} V_z, \tag{12}$$

### 5. Determining the transformation coefficients of the spatial coordinates, time and the relative speeds of the frames in the process of transition from frame K′ to frame K and the reverse transition

Let us assume that the velocity of a light ray, propagating in the proper space of frame $K'$, is $\vec{c}$ and the direction of propagation is in the plane $O'X'Y'$ at an acute angle of $\vartheta'$ to axis $O'X'$. The components of the velocity of light $\vec{c}$ along the axes $O'X'$ and $O'Y'$ will be equal to

$$c'_x = c \cos \vartheta', \quad c'_y = c \sin \vartheta'. \tag{13}$$

Let the velocity value of the same light ray relative to frame $K$ be $c_a$ and its components on the axes $OX$ and $OY$ be $c_{ax}$ and $c_{ay}$. The value of $c_a$ is equal to

$$c_a = \sqrt{c^2_{ax} + c^2_{ay}}. \tag{14}$$

On the grounds of the law of velocity summation

$$c_{ax} = \frac{v}{v'}(c.\cos \vartheta' + v'), \quad c_{ay} = \frac{v}{v'}.c.\sin \vartheta' \tag{15}$$

the following formula is derived

$$c_a = \frac{v}{v'} \sqrt{c^2 + 2cv' \cos \vartheta' + v'^2}. \tag{16}$$

Obviously, the speed $v'$ can depend only on $v$ and $c$ and not on the angle $\vartheta\,'$. To obtain the dependence needed, in the above formula $c_a$ should be replaced by $c$ and $\vartheta\,'$ by $\frac{\pi}{2}$. As a result, the second of the formulas (21) will be obtained.

It is clear from the above formula that the following formulae will be valid for the case discussed
$$v^2 dt^2 + dy^2 = c^2 dt^2, \tag{17}$$
$$dy'^2 = c^2 dt'^2. \tag{18}$$

Taking into account that the result of the differentiation of (5) and (9) is
$$dy = dy', \quad dt = \frac{v'}{v} dt' \tag{19}$$

formula (17) will look in the following way
$$dy'^2 = (c^2 - v^2)\left(\frac{v'}{v}\right)^2 dt'^2. \tag{20}$$

Since the ratio $\frac{v'}{v}$ can only be positive, formulas (18) and (20) will result in
$$\frac{v'}{v} = \frac{1}{\sqrt{1 - \frac{v^2}{c^2}}}, \quad \frac{v}{v'} = \frac{1}{\sqrt{1 + \frac{v'^2}{c^2}}}. \tag{21}$$

On the grounds of the above results, for the transformations (5) and (8) we obtain
$$x = x' + v't', \; y = y', \; z = z', \; t = t'\sqrt{1 + \frac{v'^2}{c^2}}, \; v = \frac{v'}{\sqrt{1 + \frac{v'^2}{c^2}}}. \tag{22}$$

The corresponding transformation in the transition from frame $K$ to frame $K'$ will be
$$x' = x - vt, \; y' = y, \; z' = z, \; t' = t\sqrt{1 - \frac{v^2}{c^2}}, \; v' = \frac{v}{\sqrt{1 - \frac{v^2}{c^2}}}. \tag{23}$$

### 6. Final form of the law of velocity summation

The last of the formulas (22) makes it possible to obtain the following final form of the law of velocity summation in the transition from frame $K'$ to frame $K$
$$V_x = \frac{(V_x' + v')}{\sqrt{1 + \frac{v'^2}{c^2}}}, \; V_y = \frac{V_y'}{\sqrt{1 + \frac{v'^2}{c^2}}}, \; V_z = \frac{V_z'}{\sqrt{1 + \frac{v'^2}{c^2}}}. \tag{24}$$

Having in mind the last of the formulas (23) at the reverse transition from frame $K$ to frame $K'$, the law of velocity summation can be expressed as:
$$V_x' = \frac{V_x - v}{\sqrt{1 - \frac{v^2}{c^2}}}, \; V_y' = \frac{V_y}{\sqrt{1 - \frac{v^2}{c^2}}}, \; V_z' = \frac{V_z}{\sqrt{1 - \frac{v^2}{c^2}}}. \tag{25}$$

### 7. Thomas precession

As is known (see [5]), if certain frame $K'$ rotates relative to the origin $O$ of frame $K$, the velocity of the change with time of an random vector $\vec{G}$ in this frame will satisfy the equation
$$\frac{d\vec{G}}{dt'} = \frac{d\vec{G}}{dt} - \vec{\omega}_T \times \vec{G}, \tag{26}$$
where $\vec{\omega}_T$ is the angular velocity of rotation, obtained by Thomas.

If a material body is moving along a trajectory of a rosette type there will be a precession of the perihelia and the velocity of the body on passing, will always be perpendicular to the perihelia. The direct result is that the "precession" of the velocity equals to the precession of the perihelia.

In Sec. 5.1 from [10] it is proved that with a properly chosen interval of time $\Delta t$ relative to frame $K$, the passing of time in an accelerated reference system can be assumed with the necessary research accuracy to be identical to the passing of time in the same interval in an inertial reference system. That appeared to be possible if at the beginning of the interval $\Delta t$, the speeds of accelerated and inertial reference systems are identical relative to frame $K$. It can be easily seen, that the identical conditions mentioned at the beginning of the interval $\Delta t$ are valid also at any circular movement with a constant linear speed v of the material point $P$ compared to its free movement with velocity v. Such a phenomenon has been experimentally established in [9]. According to the experimental results, obtained there, the rate of decay of a muon, freely moving along a circular orbit with a linear speed v in a strong magnetic field coincides with the rate of decay of the freely moving with the same speed muon with accuracy to 2%. This fact indicates, that at accelerations of frame $K'$ relative to frame $K$, which do not destroy frame $K'$, the last one of the formulas (10) retains its form.

Let us determine the "precession" of the velocity $\vec{V}$ when the motion is accomplished in plane $OXZ$ of frame $K$ and the vector $\vec{\omega}_T$ is perpendicular to the same plane. For this purpose in the above equation we put

$$\vec{G} = \vec{V}, \quad \frac{d\vec{V}}{dt'} = \frac{d\vec{V}}{dt}\frac{dt}{dt'}. \tag{27}$$

The result obtained will be

$$\vec{\omega}_T \times \vec{V} = \frac{d\vec{V}}{dt}\left(1 - \frac{dt}{dt'}\right) = \frac{d\vec{V}}{dt}\left(1 - \frac{1}{\sqrt{1 - \frac{V^2}{c^2}}}\right) \approx \frac{V^2}{2c^2}\frac{d\vec{V}}{dt}. \tag{28}$$

But on the grounds of the assumptions made

$$|\vec{\omega}_T \times \vec{V}| = \omega_T V \tag{29}$$

then replacing it in formula (28), the familiar formula for $\omega_T$ we obtain

$$\omega_T = \frac{Va}{2c^2}, \quad \left(a = \frac{dV}{dt}\right). \tag{30}$$

### 8. Propagation of light in optically denser moving medium

Let frame $K'$ be moving with speed v relative to frame $K$ in the standard way described in Sec.1 and the proper space $V_{k'}$ of frame $K'$ is occupied by a medium homogeneous relative to the propagation of light. Its refraction index is $n$. Let us consider that the propagation of light in $V_{k'}$ with a speed equal to $\frac{c}{n}$ is equivalent to the motion of a material body $A$, moving with the same speed in the vacuumed space $V_{k'}$. The coordinate system $K''$ with an origin $O''$ is immovably connected to the material body $A$. The axes of frame $K''$ will be assumed to be parallel and with the same direction as the respective axes of frame $K'$. Frame $K$ is moving relative to frame $K'$ with a velocity of

$$\vec{v}' = -v'\vec{i}, \tag{31}$$

The value of $v'$ can be calculated using the last of the formulas (22). The velocity of frame $K''$ relative to frame $K'$ equals to

$$\frac{\vec{c}}{n} = \frac{c}{n}\cos\psi'\,\vec{i}' + \frac{c}{n}\sin\psi'\,\vec{j}', \tag{32}$$

where $\psi'$ is the angle between the axis $O'X'$ and the velocity $\frac{\vec{c}}{n}$. Frame $K'$ is situated in the proper space $V_k$ of frame $K$. On the grounds of the last of the formulas (23), the speed $U''$ of frame $K'$ relative to frame $K''$ will be equal to

$$U'' = \frac{\frac{c}{n}}{\sqrt{1 - \frac{1}{c^2}\left(\frac{c}{n}\right)^2}} = \frac{\frac{c}{n}}{\sqrt{1 - \frac{1}{n^2}}}. \tag{33}$$

On the grounds of the above formulas and law of velocity summation in the general case, given in [10], the components of velocity $\vec{U}''_k$ of frame $K$ relative to frame $K''$, are equal to

$$U''_{xk} = \frac{-v' - \frac{c}{n}\cos\psi'}{\sqrt{1-\frac{1}{n^2}}}, \quad U''_{yk} = \frac{-\frac{c}{n}\sin\psi'}{\sqrt{1-\frac{1}{n^2}}}. \tag{34}$$

From these formulae, for speed $\vec{U}''_k$ within the order of $\frac{v'}{c}$, the following formula is obtained

$$U''_k = \frac{\sqrt{\frac{c^2}{n^2}+\frac{2cv'}{n}\cos\psi'+v'^2}}{\sqrt{1-\frac{1}{n^2}}} \approx \frac{\frac{c}{n}+v'\cos\psi'}{\sqrt{1-\frac{1}{n^2}}}. \tag{35}$$

The constant speed $V$ of a certain material body, moving along the straight trajectory $OO''$ in frame $K$ differs from the constant speed $V'$ of the same motion in frame $K'$. This difference is with a magnitude of the order of $\frac{v^2}{c^2}$, because

$$V = \frac{OO''}{t}, \quad V' = \frac{OO''}{t'}, \quad \left(t' = t\sqrt{1-\frac{v^2}{c^2}}\right), \tag{36}$$

where $t$ and $t'$ are the periods of time, needed by the body to pass the distance $OO''$ in frames $K$ and $K'$. This permits when the body is moving in the way described up to now in $V_{k'}$ and along $OO''$ trajectory, with an accuracy of the quantity of order $\frac{v}{c}$ to consider, that the motion is accomplished in the proper space of frame $K$. Therefore, the value of speed $V$ in relation to frame $K$ can be determined with the above-mentioned accuracy by the use of the last of the formulas (22). In accordance with the notations accepted, speed V becomes in the form

$$V = \frac{U''_k}{\sqrt{1+\frac{U''^2_k}{c^2}}}. \tag{37}$$

After replacing $U''_k$ by the value, obtained in formula (35) and accomplishing elementary algebraic transformations for speed $V$ in the order of $\frac{v}{c}$, the Fizeau formula is obtained

$$V \approx \frac{c}{n}+v\left(1-\frac{1}{n^2}\right)\cos\psi'. \tag{38}$$

Fizeau [6] was the first to test this formula experimentally by observing the displacement of the interference picture when the propagation of light is accomplished in a tube full of quickly running water. Later Michelson and Morley [7] tested it as well.

### 9. Proper time

A proper time of a certain inertial reference system is considered to be the time read through static synchronized clocks in the same frame.

Let the origin $O'$ of the initial frame $K'$ and the clock, connected immovably to it, pass the distance $\sqrt{dx^2+dy^2+dz^2}$ for the elementary interval of time $dt$. Since the watch mentioned is static relative to frame $K'$, then

$$dx' = dy' = dz' = 0, \tag{39}$$

the space-time intervals $ds^2$ and $ds'^2$ can be expressed in the following way

$$ds^2 = dt^2\left(c^2 - \frac{dx^2}{dt^2}-\frac{dy^2}{dt^2}-\frac{dz^2}{dt^2}\right) = c^2dt^2\left(1-\frac{v^2}{c^2}\right), \tag{40}$$

$$ds'^2 = c^2dt'^2. \tag{41}$$

Taking into account the fourth of the formulas (23) it is obvious that

$$ds^2 = ds'^2 \tag{42}$$

and the elementary time interval of the proper time $d\tau'$ in frame $K'$ is equal to

$$d\tau' = dt' = dt\sqrt{1-\frac{v^2}{c^2}}. \tag{43}$$

Since in the case discussed the interval $ds'$ is invariant it is obvious that the proper time $d\tau'$ will be an invariant as well, because

$$d\tau' = \frac{1}{c} ds'. \tag{44}$$

### 10. Four-dimensional vectors

As is known (see [8]) the magnitudes $x$, $y$, $z$ and $ict$ can be presented as components of a "four-dimensional radius vector" in a four-dimensional space. Using

$$x_1 = x, \; x_2 = y, \; x_3 = z, \; x_4 = ict, \; \beta = \frac{v}{c} \tag{45}$$

the components of the four-dimensional radius vector of the transition of frame $K'$ to frame $K$ will transform in accordance with (22), using the formulas

$$x_1 = x_1' - \frac{i\beta \, x_4'}{\sqrt{1-\beta^2}}, \; x_2 = x_2', \; x_3 = x_3', \; x_4 = \frac{x_4'}{\sqrt{1-\beta^2}}. \tag{46}$$

Taking into account (23), the transformation in the transition of frame $K$ to frame $K'$ will be presented as

$$x_1' = x_1 + i\beta \, x_4, \; x_2' = x_2, \; x_3' = x_3, \; x_4' = x_4 \sqrt{1-\beta^2}. \tag{47}$$

We introduce a four-dimensional vector $A_i$ with components $A_1$, $A_2$, $A_3$, $A_4$. Under the transformation of the four-dimensional coordinate system, they will transform like the components $x_i$ or $x_i'$ $(i = 1, 2, 3, 4)$. In the transition of frame $K'$ to frame $K$ the components of the four-dimensional vector $A_i$ will transform thus

$$A_1 = A_1' - \frac{i\beta \, A_4'}{\sqrt{1-\beta^2}}, \; A_2 = A_2', \; A_3 = A_3', \; A_4 = \frac{A_4'}{\sqrt{1-\beta^2}}. \tag{48}$$

The formulas for the reverse transition will be

$$A_1' = A_1 + i\beta \, A_4, \; A_2' = A_2, \; A_3' = A_3, \; A_4' = A_4 \sqrt{1-\beta^2}. \tag{49}$$

### 11. Four-dimensional velocity and four-dimensional acceleration

The four-dimensional velocity is a vector with components

$$u_i = \frac{dx_i}{ds} \quad (i = 1, 2, 3, 4). \tag{50}$$

where for $x_i$ and $ds$ formulas (40) and (45) are in force. Using (50), we obtain

$$u_1 = \frac{dx}{cdt\sqrt{1-\beta^2}} = \frac{v_x}{c\sqrt{1-\beta^2}}, \; u_2 = \frac{v_y}{c\sqrt{1-\beta^2}}, \tag{51}$$

$$u_3 = \frac{v_z}{c\sqrt{1-\beta^2}}, \; u_4 = \frac{i}{\sqrt{1-\beta^2}}. \tag{52}$$

These formulas for the components of the four-dimensional velocity show that it is a dimensionless quantity.

Further, for brevity, on repeating one and same Latin index while multiplying two quantities, for example $a_i b_i$ $(i = 1, 2, 3, 4)$, we shall understand the summation of these products at index equal to 1, 2, 3, 4, i.e.

$$a_i b_i = \sum_{i=1}^{4} a_i b_i. \tag{53}$$

For the components of the four-dimensional velocity the following formula is in force

$$u_i^2 = -1, \tag{54}$$

hence

$$dx_i^2 = -ds^2 \tag{55}$$

therefore it is a single four-dimensional vector.

The derivative of the four-dimensional vector $s$ with components

$$\frac{d^2 x_i}{ds^2} = \frac{du_i}{ds} \quad (i = 1, 2, 3, 4) \tag{56}$$

is called a four-dimensional acceleration.

The four-dimensional velocity and acceleration are "mutually perpendicular", because the result of the differentiation of the two parts of formula (54) is

$$u_i \frac{du_i}{ds} = 0. \tag{57}$$

# DYNAMICS OF THE VACUUM FIELD THEORY

G. Gemedjiev
Plovdiv University "Paisii Hilendarski", 24 Tsar Asen Str. 4000 Plovdiv, Bulgaria
e-mail: gemedjievg@mail.bg

*In kinematics of the vacuum field theory we proved the invariance of the proper time in the inertial reference system $K'$. This is used in dynamics of the vacuum field theory to find a Langrangian of the free particle, coinciding with that in special relativity. This and the usage of the familiar methods proved that the formulas in special relativity are valid for the dynamic quantities – mass, force, momentum and energy. The investigations showed that the effective cross-section is an invariant and the formulas of particle decay are the same as those in special relativity. The formulas of the four-dimensional mode of force and impulse are derived using the formulas of the four-dimensional vector, obtained in kinematics of the vacuum field theory. The elastic collision of two particles was investigated in the general case. The difference registered between the formulas, concerning the elastic collision of two particles in the laboratory system and the center of mass system in special relativity for the particular case, in which one of the particles is immovable, was a result of the fact, that in accordance with the vacuum field theory, the treatment of the case relative to the center of mass system is devoid of physical sense. The investigation of Compton effect resulted in the well-known formula.*

The vacuum field theory [1] proved the invariance of proper time in inertial reference system $K'$. The Langrangian of the free particle identical with the one in special relativity was obtained taking into account the invariance in the method, presented in [2]. After this and the use of the methods, presented in [2], we obtain, that the formulas of special relativity will be valid for mass, force, momentum, energy, as well particle decay is found to be uniform with the one from special relativity.

1. **4-momentum and 4-force**

The method presented in [2] we come to the formula below

$$\frac{\delta S}{\delta x_i} = m_0 c u_i, \quad \left( u_i = \frac{dx_i}{\sqrt{-dx_i^2}} \right) \tag{1}$$

where $S$ is the action and $u_i$ is the four-dimensional velocity. In this formula

$$p_i = \frac{\partial S}{\partial x_i} \tag{2}$$

are components of the 4-momentum. These components for a free material particle are equal to
$$p_i = m_0 c u_i. \tag{3}$$

As is well known from mechanics, the derivatives $\frac{\partial S}{\partial x}, \frac{\partial S}{\partial y}, \frac{\partial S}{\partial z}$ are the three components of momentum of particle $\vec{p}$ and the derivative $-\frac{\partial S}{\partial t} = -ic \frac{\partial S}{\partial x_4}$ is the energy of particle $E$. The (51) and (52) formulas, presenting the four-dimensional velocity, discussed in [1] make certain that the spatial components $p_i$ actually coincide with those of momentum $\vec{p}$ and that the time component equals to $\frac{iE}{c}$, i. e.
$$p_1 = p_x, \; p_2 = p_y, \; p_3 = p_z, \; p_4 = \frac{iE}{c}. \tag{4}$$

The comparison of the latter formulas to the formulas, denoted by (45) and presented in [1], shows that momentum and energy are components of a four-dimensional radius vector. So the formulas (46), presented in [1] and concerning the transformation of the four-dimensional vector in the transition from one coordinate system to another, will be valid. Then
$$p_x = p'_x + \frac{vE'}{c^2 \sqrt{1 - \frac{v^2}{c^2}}}, \; p_y = p'_y, \; p_z = p'_z, \; E = \frac{E'}{\sqrt{1 - \frac{v^2}{c^2}}}. \tag{5}$$

The 4-force $\vec{g}$ is introduced by analogy with the ordinary three-dimensional vector of force $\vec{f}$ and its components are
$$g_i = \frac{dp_i}{ds} = m_0 c \frac{du_i}{ds}. \tag{6}$$

Taking into account that $ds = cdt\sqrt{1 - \frac{v^2}{c^2}}$, then the components $g_i$ ($i$ = 1, 2, 3) can be presented using the components $f_i$ of the ordinary three-dimensional vector $\vec{f}$ by the following formulas
$$g_i = \frac{1}{c\sqrt{1 - \frac{v^2}{c^2}}} \frac{dp_i}{dt} = \frac{f_i}{c\sqrt{1 - \frac{v^2}{c^2}}}. \tag{7}$$

We will use formula (57) from [1] to obtain the formula for $g_4$. The following equation is easily derived out of this formula
$$g_i u_i = 0. \tag{8}$$

Replacing $g_i$ and $u_i$ in the above equation by their equals and effecting the respective simplifications for $g_4$ we obtain
$$g_4 = \frac{i\vec{f}.\vec{v}}{c^2\sqrt{1 - \frac{v^2}{c^2}}}. \tag{9}$$

The following ratio is a direct result from the mode of the components of momentum (4) and formula (54) from [2]
$$p_i^2 = m_0^2 c^2. \tag{10}$$

## 2. Theory of the elastic collision between two particles in the general case

In 1914 Jüttner [3] developed the general case of the theory of the elastic collision between two particles in laboratory system in accordance with special relativity. Two basic methods are used in the treatment of particular cases of elastic collision between two particles in special relativity when one of them is supposed to be immovable in laboratory system. The method, presented in [2], Sec. 13, first investigates the impact relative to the center of mass system. Then, using the formulas of momentum and energy transformation in the transition from the center of mass system to laboratory system, the values of the parameters in the second system are obtained. The second method investigates the collision relative to laboratory system, treating the three separate cases, discussed in [3], [2], Sec. 13, [4] and [5]. Based on the vacuum field theory this work presents a theory of the elastic collision between two particles in laboratory system for the general case that differs from the theory of Jüttner. This theory makes it possible to obtain the above-mentioned particular cases as done in [6].

The position of particles 1 and 2 before and after the impact is presented in Fig. 1. A second before the collision they are positioned in points $A_1$ and $A_2$. Their velocities are $\vec{V}_{1,0}$ and $\vec{V}_{2,0}$, respectively. The moving particles are connected to the right-orientated coordinate system $O''X''Y''Z''$ in such a way that the velocity vectors $\vec{V}_{1,0}$ and $\vec{V}_{2,0}$ should be situated in plane $O''X''Z''$ and the axis $O''X''$ should be parallel to the straight line $\overline{A_1 A_2}$. The point of the collision of the particles $O'$ serves as origin of the coordinate system $O'X'Y'Z'$, immovable in laboratory system. Its plane $O'X'Z'$ coincides with plane $O''X''Z''$. The axes of the system $O'X'Y'Z'$ are parallel to the respective axes of system $O''X''Y''Z''$.

On the grounds of the above stated, the single vectors $\vec{i}''$, $\vec{j}''$ and $\vec{k}''$ along the axes $O''X''$, $O''Y''$ and $O''Z''$ will be equal to

$$\vec{i}'' = \frac{\vec{V}_{1,0} - \vec{V}_{2,0}}{|\vec{V}_{1,0} - \vec{V}_{2,0}|}, \quad \vec{j}'' = \frac{\vec{V}_{2,0} \times \vec{V}_{1,0}}{|\vec{V}_{2,0} \times \vec{V}_{1,0}|}, \quad \vec{z}'' = \vec{i}'' \times \vec{j}''. \tag{11}$$

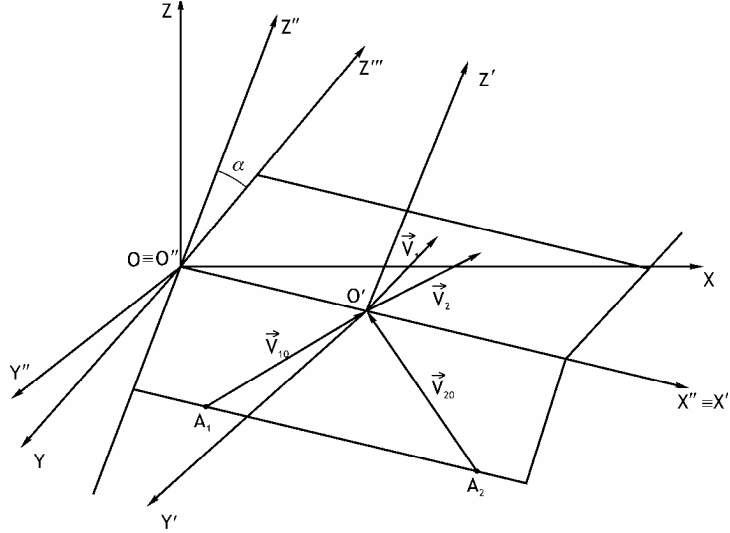

**Fig. 1.**

The motion of particles 1 and 2 before the collision can be expanded into a motion along the axis $O''X''$ and a motion of the axis itself in plane $O'X'Z'$. The axes $O''X''$ and $O''Z''$ always remain parallel to axes $OX'$ and $O'Z'$. The velocity of the motion of axis $O''X''$ relative to axis $OX'$ is

$$V_z'' = V_{1,0}\sin\vartheta_1 = V_{1,0}\frac{\left|(\vec{V}_{1,0} - \vec{V}_{2,0}) \times \vec{V}_{1,0}\right|}{|\vec{V}_{1,0}| \cdot |\vec{V}_{1,0} - \vec{V}_{2,0}|} = \frac{|\vec{V}_{1,0} \times \vec{V}_{2,0}|}{|\vec{V}_{1,0} - \vec{V}_{2,0}|}. \tag{12}$$

where $\vartheta_1$ is the angle between the velocity $\vec{V}_{1,0}$ and the axis $O''X''$.

The components of velocities $V_{1x,0}$ and $V_{2x,0}$ of particles 1 and 2 along the axis $O''X''$ before their collision are

$$V_{1x,0} = V_{1,0}\cos\vartheta_1 = V_{1,0}\frac{\left|(\vec{V}_{1,0} - \vec{V}_{2,0}) \cdot \vec{V}_{1,0}\right|}{|\vec{V}_{1,0}| \cdot |\vec{V}_{1,0} - \vec{V}_{2,0}|} = \frac{\left|(\vec{V}_{1,0} - \vec{V}_{2,0}) \cdot \vec{V}_{1,0}\right|}{|\vec{V}_{1,0} - \vec{V}_{2,0}|}, \tag{13}$$

$$V_{2x,0} = V_{2,0}\cos\vartheta_2 = V_{2,0}\frac{\left|(\vec{V}_{1,0} - \vec{V}_{2,0}) \cdot \vec{V}_{2,0}\right|}{|\vec{V}_{2,0}| \cdot |\vec{V}_{1,0} - \vec{V}_{2,0}|} = \frac{\left|(\vec{V}_{1,0} - \vec{V}_{2,0}) \cdot \vec{V}_{2,0}\right|}{|\vec{V}_{1,0} - \vec{V}_{2,0}|}. \tag{14}$$

where $\vartheta_2$ is the angle between the velocity $\vec{V}_{2,0}$ and the axis $O''X''$.

In accordance with the law of conservation of momentum it follows that, after the collision, the two particles remain in plane $O''X''Z'''$. This plane is at an angle $\alpha$ with plane $O''X'Z''$. In accordance with special relativity and the vacuum field theory the energies and the momentum components of particles 1 and 2, before the collision in $O'X'Y'Z'$ coordinate system, are respectively equal to

$$E_{1,0} = \frac{m_1 c^2}{\sqrt{1 - \frac{V_{1,0}^2}{c^2}}}, \quad p_{1x,0} = \frac{m_1 V_{1x,0}}{\sqrt{1 - \frac{V_{1,0}^2}{c^2}}}, \quad p_{1y,0} = 0, \quad p_{1z,0} = \frac{m_1 V_z''}{\sqrt{1 - \frac{V_{1,0}^2}{c^2}}}, \tag{15}$$

$$E_{2,0} = \frac{m_2 c^2}{\sqrt{1-\frac{V_{2,0}^2}{c^2}}}, \quad p_{2x,0} = \frac{m_2 V_{2x,0}}{\sqrt{1-\frac{V_{2,0}^2}{c^2}}}, \quad p_{2y,0} = 0, \quad p_{2z,0} = \frac{m_2 V_z''}{\sqrt{1-\frac{V_{2,0}^2}{c^2}}}. \tag{16}$$

The values of these quantities after the collision as well as the components of the velocity will be denoted without the index 0.

The laws of conservation of momentum and energies of the particles will be presented in the following way:

$$p_{1i} + p_{2i} = p_{1i,0} + p_{2i,0}, \quad (i = x, y, z), \tag{17}$$

$$E_1 + E_2 = E_{1,0} + E_{2,0}. \tag{18}$$

With the reservation concerning the orientation of plane $O''X''Z'''$, the velocities $\vec{V}_1'$ and $\vec{V}_2'$ of the particles in $O'X'Y'Z'$ coordinate system can be presented as follows

$$\vec{V}_1' = V_1 \cos \iota_1 \vec{i}' + V_1 \sin \iota_1 \sin \alpha \vec{j}' + (V_z'' + V_1 \sin \iota_1 \cos \alpha) \vec{k}',$$

$$\vec{V}_2' = V_2 \cos \iota_2 \vec{i}' + V_2 \sin \iota_2 \sin \alpha \vec{j}' + (V_z'' + V_2 \sin \iota_2 \cos \alpha) \vec{k}', \tag{19}$$

where $\vec{i}'$, $\vec{j}'$ and $\vec{k}$ are the unit vectors of the axes $O'X'$, $O'Y'$ and $O'Z'$. This makes it possible to present (17) and (18) formulas in the following way

$$\frac{m_1 V_1 \cos \iota_1}{\sqrt{1-\frac{V_1^2}{c^2}}} + \frac{m_2 V_2 \cos \iota_2}{\sqrt{1-\frac{V_2^2}{c^2}}} = p_{1x,0} + p_{2x,0}, \tag{20}$$

$$\frac{m_1 V_1 \sin \iota_1 \sin \alpha}{\sqrt{1-\frac{V_1^2}{c^2}}} - \frac{m_2 V_2 \sin \iota_2 \sin \alpha}{\sqrt{1-\frac{V_2^2}{c^2}}} = 0, \tag{21}$$

$$\frac{m_1(V_z'' + V_1 \sin \iota_1 \cos \alpha)}{\sqrt{1-\frac{V_1^2}{c^2}}} + \frac{m_2(V_z'' - V_2 \sin \iota_2 \cos \alpha)}{\sqrt{1-\frac{V_2^2}{c^2}}} = \frac{m_1 V_z''}{\sqrt{1-\frac{V_{1,0}^2}{c^2}}} + \frac{m_2 V_z''}{\sqrt{1-\frac{V_{2,0}^2}{c^2}}}, \tag{22}$$

$$\frac{m_1 c^2}{\sqrt{1-\frac{V_1^2}{c^2}}} + \frac{m_2 c^2}{\sqrt{1-\frac{V_2^2}{c^2}}} = \frac{m_1 c^2}{\sqrt{1-\frac{V_{1,0}^2}{c^2}}} + \frac{m_2 c^2}{\sqrt{1-\frac{V_{2,0}^2}{c^2}}}. \tag{23}$$

It is easy to establish that the latter three of the above equations are compatible.

In case of certain meanings of the quantities, characterizing the motion of the two particles before their collision, equation (20) and the latter two equations form a system of three equations for the five unknown parameters $V_1$, $\iota_1$, $V_2$, $\iota_2$ and $\alpha$. While determining the first three parameters as functions of $\iota_2$ and $\alpha$, the following formulas are obtained

$$(V_2)_{1,2} = \frac{\beta_2 \gamma + a_2^2 \delta_2 \pm a_2 \sqrt{a_2^2(\delta_2 + \varepsilon) + 2\beta_2 \gamma \delta_2 + \beta_2^2 \varepsilon - \gamma^2}}{\beta_2^2 + a_2^2} \tag{24}$$

where

$$a_2 = \frac{\frac{E_0^2}{c^2} - p_{x0}^2 - p_{z0}^2 - m_2^2 c^2 + m_1^2 c^2}{2 m_2}, \tag{25}$$

$$\beta_2 = p_{x0} \cos \iota_2 + p_{z0} \sin \iota_2 \cos \alpha, \quad \gamma = E_0 + p_{z0} V_z'', \tag{26}$$

$$\delta_2 = V_z'' \sin \iota_2 \cos \alpha, \quad \varepsilon = 1 - V_z''^2. \tag{27}$$

Let us again denote the radical, corresponding to the case we discuss, by $V_2$. Then

$$\iota_1 = \text{arcctg}\left(\frac{p_{x0}\sqrt{1-\frac{V_2'^2}{c^2}}}{m_2 V_2 \sin \iota_2} - \cot \iota_2\right), \tag{28}$$

$$V_1 = \frac{p_{x0} - \dfrac{m_2 V_2 \cos \iota_2}{\sqrt{1 - \dfrac{V_2'^2}{c^2}}}}{\left(\dfrac{E_0}{c^2} - \dfrac{m_2}{\sqrt{1 - \dfrac{V_2'^2}{c^2}}}\right) \cos \iota_1}. \qquad (29)$$

The formulas (20), (22) and (25) concerning the energy $E_2$ of the particle after the collision lead to the following formula

$$E_2 = \frac{a_2 m_2 c^2}{E_0 - p_{x0} V_2 \cos \iota_2 - p_{z0}(V_2 \sin \iota_2 \cos \alpha - V_z'')}. \qquad (30)$$

The participation of $V_1$, $\iota_1$, $V_2$, $\iota_2$ in (20) – (23) formulas is symmetrical. To find the functional dependence of $V_1$ and $E_1$ relative to $\iota_1$ it would only be necessary to change the places of indices 1 and 2 in all the parameters used to determine the functional dependence of $V_2$ and $E_2$ relative to $\iota_2$.

### 3. The treatment of the elastic collision of two particles in the center of mass system is devoid of physical sense

Let us consider a particular case in special relativity when one of the particles is immovable in laboratory system again. It is possible initially to treat the elastic collision in the center of mass system and then using the familiar transformations of momentums and energies of the particles in the transition from the center of mass system to laboratory system to find the parameters of their motion in the latter system. This case is discussed in [2] p. 49 and the following formulas for the energies of the particles after their collision in laboratory system are obtained

$$E_1' = E_{1,0} - \frac{m_2 c^2 p_0^2}{(m_1 + m_2)c^2 + 2m_2 E_{1,0}}(1 - \cos \iota_2'), \qquad (31)$$

$$E_2' = E_{2,0} + \frac{m_2 c^2 p_0^2}{(m_1 + m_2)c^2 + 2m_2 E_{1,0}}(1 - \cos \iota_2'), \qquad (32)$$

where $\iota_2'$ is the angle of scattering of the second particle in the center of mass system.

As is known [5] p. 135 the angle $\iota_2'$ connects with the angle of scattering of the second particle $\iota_{2L}$ in laboratory system by the following dependence

$$tg\, \iota_{2L} = \frac{tg\, \iota_2'}{\sqrt{1 - \dfrac{V^2}{c^2}}}, \qquad (33)$$

where the speed $V$ of center of mass system relative to laboratory system in accordance with [1] p. 48 is equal to

$$V = \frac{p_0 c^2}{E_0}. \qquad (34)$$

The trigonometric and algebraic transformations of $\cos \iota_2'$ result in

$$\cos \iota_2' = \frac{E_0 \cos \iota_{2L}}{\sqrt{E_0^2 - p_0^2 c^2 \sin^2 \iota_{2L}}} \qquad (35)$$

and consequently formulas (31) and (32) can be expressed in the following way

$$E_1' = E_{1,0} - \frac{m_2 c^2 p_0^2}{(m_1^2 + m_2^2)c^2 + 2m_2 E_{1,0}}\left(1 - \frac{E_0 \cos \iota_{2L}}{\sqrt{E_0^2 - p_0^2 c^2 \sin^2 \iota_{2L}}}\right), \qquad (36)$$

$$E_2' = E_{2,0} + \frac{m_2 c^2 p_0^2}{(m_1^2 + m_2^2)c^2 + 2m_2 E_{1,0}}\left(1 - \frac{E_0 \cos \iota_{2L}}{\sqrt{E_0^2 - p_0^2 c^2 \sin^2 \iota_{2L}}}\right). \qquad (37)$$

The investigation shows that the analytical form of the latter two formulas cannot be reduced to these formulas from the corresponding particular case from the discussed general case in Sec. 2.7 by [6]. The only exception is in the case of $\iota_{2L} = \pi$. In this case, taking into account the minus in front of the square root of formula (2.109) from [6], the two methods give the same analytical expressions for $E_1'$ and $E_2'$

$$E_1' = \frac{E_{1,0}c^2\left(m_1^2 + m_2^2\right) + 2m_1^2 m_2 c^4}{\left(m_1^2 + m_2^2\right)c^2 + 2m_2 E_{1,0}}, \tag{38}$$

$$E_2' = \frac{m_2\left(2E_{1,0}^2 + 2E_{1,0}m_2 c^2 + \left(m_2^2 - m_1^2\right)c^4\right)}{\left(m_1^2 + m_2^2\right)c^2 + 2m_2 E_{1,0}}. \tag{39}$$

The conclusion is that in special relativity when applying the latter two methods, there is not any simple definition of the connection between the physical quantities. The investigations proved that the differences between the two methods increase with the increase of the velocity of the first body, before the collision, to values comparable to the velocity of light. For example, if $m_1 = 2m_2$, $V_{10} = 0{,}8c$, $\iota_1 = 30^0$, the value of $E_1'$ obtained, using the second method is smaller by 17,0% than the one calculated using the first method. At the same values of ratio of masses, $V_{10}$ and $\iota_1$ the value for $E_2'$ using the second method is greater by 47,0% than the one obtained using the first method. This difficulty in special relativity is surmounted in the vacuum field theory, on the grounds of which only the first method is true. In accordance with the vacuum field theory the second method is devoid of physical sense due to the fact that the energies and the momentum of the particles are determined only relative to the proper space domain of the experimental installation and not related with the proper space of the center of mass system because it does not possess one. The mistakes with the second method are also due to the space indeterminacy of the phenomena discussed and, in the particular case, instead of formula (33) in the vacuum field theory the formula $tg\iota_{2L} = tg\iota_2'$ is used.

## 4. Compton effect

The quantum theory considers the light wave as a flux of light quanta – photons. The Compton effect is treated as an elastic collision of two particles - the photon moving with the speed of light and the immovable in laboratory system electron whose mass is $m_2$. Assuming that the direction of the distribution of the light wave coincides with the positive direction of the axis $OX$ and using the designations for energy $E_{1,0}$ and the component of momentum $p_{1x,0}$ of the photon before the collision, accepted in the previous section, the formulas look as follows can be written in the form

$$E_{1,0} = h\nu_0, \tag{40}$$

$$p_{1x,0} = p_0 = \frac{h\nu_0}{c}, \tag{41}$$

where $\nu_0$ is the frequency of the of the falling light. Fig. 2 presents a scheme of the collision between the photon and the electron. In the same figure we denote by $\vec{p}_1$ and $\vec{p}_2$ the moments of the photon and the electron after the collision, and $\vartheta$ and $\varphi$ are the angles between these moments and the direction of light propagation. The law of energy conservation and the expression for $m_1 a_1''$ where $a_1''$ satisfies formula (2.104) in [6], are valid for laboratory system. In the case discussed, they are expressed as follows

$$E_0 = h\nu_0 + m_2 c^2, \tag{42}$$

$$m_1 a_1'' = m_2 h\nu_0 + m_1^2 c^2. \tag{43}$$

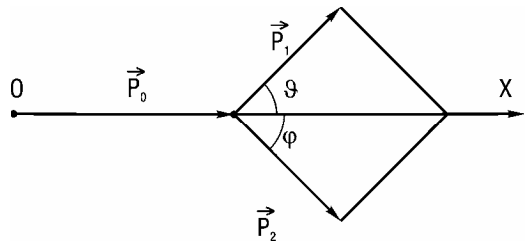

Fig. 2.

Let us use formula (2.109) from [6] to find the connection between the frequencies $\nu_0$ and $\nu$ of the photon before and after the collision, respectively, and the angle $\vartheta$. Taking into account the above formulas and the fact, that the rest mass of the photon $m_1$ is equal to zero and its energy after the collision is $h\nu$, the following formula is obtained

$$h\nu = \frac{m_2 h\nu_0(h\nu_0 + m_2 c^2) + m_2\left(\dfrac{h\nu_0}{c}\right)^2 c^2 \cos\vartheta}{\dfrac{(h\nu_0 + m_2 c^2)^2}{c^2} - \left(\dfrac{h\nu_0}{c}\right)^2 \cos^2\vartheta}, \tag{44}$$

hence

$$\nu = \nu_0 \frac{m_2 c^2}{m_2 c^2 + h\nu_0 (1-\cos\vartheta)}. \tag{45}$$

As is known, the wavelengths of the photon before and after the collision with the electron are equal to

$$\lambda_0 = \frac{c}{\nu_0}, \tag{46}$$

$$\lambda = \frac{c}{\nu}. \tag{47}$$

Using formula (45) we express the connection between them this way

$$\lambda = \lambda_0 + \lambda_c (1-\cos\vartheta), \tag{48}$$

where $\lambda_c = \frac{h}{mc}$ is called Compton wavelength of the electron. As it is seen by the formula, the change of wavelength of light after the collision is equal to $\lambda_c (1-\cos\vartheta)$.

### 5. Invariance of the effective cross-section

Let us consider the collision of a beam of particle with density $n_2$ (i.e. the number of particles in a volume unit) with the particles of a target, immovable relative to frame $K$, with density $n_1$. Let us assume that the beam particles are immovable relative to frame $K'$, which moves relative to frame $K$ in the standard way, given in [1] Sec. 1 with speed v.

As is known, the number of collisions $d\nu$ in the elementary volume $dV$ of frame $K$ is

$$d\nu = \sigma \mathrm{v} n_1 n_2 dV dt, \tag{49}$$

where the coefficient $\sigma$ is the effective cross section in frame $K$.

The number of collision $d\nu'$ in frame $K'$ is equal to

$$d\nu' = \sigma' \mathrm{v}' n_1' n_2' dV' dt', \tag{50}$$

where all the prime quantities are relative to frame $K'$.

In their essence the number of collision $d\nu$ is an invariant value. The space in the vacuum field theory is a Newtonean space, so the volumes are invariant as well. This is valid both for the unit and the elementary volumes. Consequently,

$$d\nu = \mathrm{inv}, \; dV = dV', \; n_1 = n_1', \; n_2 = n_2'. \tag{51}$$

Taking into account that in accordance with formula (10) from [2] we have

$$\mathrm{v} dt = \mathrm{v}' dt', \tag{52}$$

the effective cross section $\sigma$ appears to be an invariant as well, i. e.

$$\sigma = \mathrm{inv}. \tag{53}$$

# VACUUM FIELD THEORY - ELECTRODYNAMICS OF INERTIAL REFERENCE SYSTEM MOVING IN VACUUM

**G. Gemedjiev**

Plovdiv University "Paisii Hilendarski", 24 Tsar Asen Str. 4000 Plovdiv, Bulgaria
e-mail: gemedjievg@mail.bg

*This publication proves the covariance of Maxwell equations in the transition from the reference system $K$, connected with vacuum in a limited domain to the inertial reference system $K'$, moving relative to it in the same spatial domain. To prove*

*this in the vacuum field theory, the transformation of the coordinates, of time and relative velocities of the above-mentioned transition are used. The two new functions introduced are the same as in special relativity. The consideration of Doppler effect in the vacuum field theory results in the derivation of its well-known formula in the relative case. The real and observed aberrations, as well the reflection from a moving mirror are investigated. As a result, the formulas for them as well as the change of amplitude of the electromagnetic wave were derived. The differences between these formulas and the corresponding relative formulas, occurring in some cases, appear to be insignificant and experimentally unobservable. The well-known formula from the experiment, carried out to determine the light pressure, was obtained in the process of its consideration in the vacuum field theory. The question concerning the velocity of motion of the inertial reference system $K''$ in the proper space of the inertial reference system $K'$, in case its mass and dimensions are much smaller than those of $K'$, is considered theoretically at the end. On the grounds of the results obtained an experimental check can be accomplished nowadays.*

## 1. Covariance of electrodynamics equations

Let us assume that the inertial reference system $K'$ and the immovable substance, occupying its space move with a constant velocity $\vec{v}$ relative to frame $K$ in the standard way, presented in Sec. 1 of [1]. In the consideration of the motion of frame $K'$ relative to frame $K$ the real coordinates $x_1 = x$, $x_2 = y$, $x_3 = z$, and $x_4 = ct$ will be used. The transformation (23) from [1] will be expressed as follows

$$x'_1 = x_1 - \beta x_4,\ x'_2 = x_2,\ x'_3 = x_3,\ x'_4 = \gamma x_4, \tag{1}$$

where

$$\beta = \frac{v}{c},\ \gamma = \sqrt{1 - \frac{v^2}{c^2}}. \tag{2}$$

The Maxwell equations for the immovable substance in frame $K$ will be of the form

$$rot\vec{H} - \frac{1}{c}\frac{\partial \vec{D}}{\partial t} = \frac{4\pi \vec{I}}{c}, \tag{3}$$

$$rot\vec{E} - \frac{1}{c}\frac{\partial \vec{B}}{\partial t} = 0, \tag{4}$$

$$div\vec{D} = 4\pi \rho, \tag{5}$$

$$div\vec{B} = 0. \tag{6}$$

Substituting in the above equations

$$\partial_p = \frac{\partial}{\partial x_p},\ (p = 1, 2, 3),\ \partial_4 = \frac{\partial}{\partial x_4} = \frac{1}{c}\frac{\partial}{\partial t}, \tag{7}$$

and introducing the four-dimensional vector of the current density $\vec{J}$ with components

$$J_i = -\frac{4\pi I_i}{c},\ (i = 1, 2, 3),\ J_4 = 4\pi \rho \tag{8}$$

the equations will be of the form

$$\partial_p H_q - \partial_q H_p - \partial_4 D_r = -J_r, \tag{9}$$

$$\partial_p E_q - \partial_q E_p - \partial_4 B_r = 0, \tag{10}$$

$$\sum_{p=1}^{3} \partial_p D_p = J_4, \tag{11}$$

$$\sum_{p=1}^{3} \partial_p B_p = 0, \tag{12}$$

where $p, q, r = 1, 2, 3$ and the indices $p, q, r$ in formulas (9) and (10) form a cyclic permutation of the figures 1, 2, 3.

We shall prove that Maxwell equations preserve their forms in frame $K'$, i.e.

$$\partial'_p H'_q - \partial'_q H'_p - \partial'_4 D'_r = -J'_r, \tag{13}$$

$$\sum_{p=1}^{3} \partial'_p D'_p = J'_4, \tag{14}$$

$$\partial'_p E'_q - \partial'_q E'_p - \partial'_4 B'_r = 0, \tag{15}$$

$$\sum_{p=1}^{3} \partial_p' B_p' = 0, \tag{16}$$

after transformation (1) is applied to them only if the components of $J(J_r, J_4)$ are transformed into components of a four-dimensional vector, i.e.

$$J_1' = J_1 + \beta J_4, \; J_2' = J_2, \; J_3' = J_3, \; J_4' = \gamma J_4, \tag{17}$$

and if the components of the vectors $\vec{H}$, $\vec{D}$, $\vec{B}$, $\vec{E}$ are transformed in the following way

$$D_1' = \gamma D_1, \; D_2' = \gamma D_2, \; D_3' = \gamma D_3, \tag{18}$$

$$H_1' = H_1, \; H_2' = H_2 + \beta D_3, \; H_3' = H_3 - \beta D_2, \tag{19}$$

$$B_1' = \gamma B_1, \; B_2' = \gamma B_2, \; B_3' = \gamma B_3, \tag{20}$$

$$E_1' = E_1, \; E_2' = E_2 - \beta B_3, \; E_3' = E_3 + \beta B_2. \tag{21}$$

In accordance with the differentiation rules the partial derivative of coordinate $x_i$, ($i = 1, 2, 3, 4$) will be equal to

$$\frac{\partial}{\partial x_i} = \sum_{k=1}^{4} \frac{\partial x_k'}{\partial x_i} \frac{\partial}{\partial x_k'}. \tag{22}$$

Taking into account transformation (1) for the partial derivatives of coordinate $x_i$, the result will be

$$\frac{\partial}{\partial x_i} = \frac{\partial}{\partial x_i'}, \; (i = 1, 2, 3), \tag{23}$$

$$\frac{\partial}{\partial x_4} = -\beta \frac{\partial}{\partial x_1'} + \gamma \frac{\partial}{\partial x_4'}. \tag{24}$$

This allows to write the equations (11) and (12) in the form

$$\sum_{k=1}^{3} \frac{\partial D_k}{\partial x_k'} = \gamma \sum_{k=1}^{3} \frac{\partial D_k'}{\partial x_k'} = J_4 = \gamma J_4', \tag{25}$$

$$\sum_{k=1}^{3} \frac{\partial B_k}{\partial x_k'} = \gamma \sum_{k=1}^{3} \frac{\partial B_k'}{\partial x_k'} = 0. \tag{26}$$

By multiplying the above two equations by $\gamma^{-1}$ we obtain that equations (14) and (16) are really valid. Equations (9) and (10) will obtain the form below after using formulas (17) – (21)

$$\frac{\partial \left( H_3' + \frac{\beta}{\gamma} D_2' \right)}{\partial x_2'} - \frac{\partial \left( H_2' - \frac{\beta}{\gamma} D_3' \right)}{\partial x_3'} - \left( -\beta \frac{\partial}{\partial x_1'} + \frac{1}{\gamma} \frac{\partial}{\partial x_4'} \right) \gamma D_1' =$$

$$= \frac{\partial H_3'}{\partial x_2'} - \frac{\partial H_2'}{\partial x_3'} - \frac{\partial D_1'}{\partial x_4'} + \frac{\beta}{\gamma} \sum_{k=1}^{3} \frac{\partial D_k'}{\partial x_k'} = -J_1' + \frac{\beta}{\gamma} J_4', \tag{27}$$

$$\frac{\partial H_1'}{\partial x_3'} - \frac{\partial \left( H_3' + \frac{\beta}{\gamma} D_2' \right)}{\partial x_1'} - \left( -\beta \frac{\partial}{\partial x_1'} + \gamma \frac{\partial}{\partial x_4'} \right) \frac{D_2'}{\gamma} = -J_2', \tag{28}$$

$$\frac{\partial \left( H_2' - \frac{\beta}{\gamma} D_3' \right)}{\partial x_1'} - \frac{\partial H_1'}{\partial x_2'} - \left( -\beta \frac{\partial}{\partial x_1'} + \gamma \frac{\partial}{\partial x_4'} \right) \frac{D_3'}{\gamma} = -J_3', \tag{29}$$

and taking into account equation (25), the validity of equation (13) becomes obvious.

Finally, by analogy, from equation (10) we shall obtain

$$\frac{\partial \left( E_3' - \frac{\beta}{\gamma} B_2' \right)}{\partial x_2'} - \frac{\partial \left( E_2' + \frac{\beta}{\gamma} B_3' \right)}{\partial x_3'} + \left( -\beta \frac{\partial}{\partial x_1'} + \gamma \frac{\partial}{\partial x_4'} \right) \frac{B_1'}{\gamma} = \frac{\partial E_3'}{\partial x_2'} - \frac{\partial E_2'}{\partial x_3'} + \frac{\partial B_1'}{\partial x_4'} - \frac{\beta}{\gamma} \sum_{k=1}^{3} \frac{\partial B_k'}{\partial x_k'} = 0, \tag{30}$$

$$\frac{\partial E_1'}{\partial x_3'} - \frac{\partial \left( E_3' - \frac{\beta}{\gamma} B_2' \right)}{\partial x_1'} + \left( -\beta \frac{\partial}{\partial x_1'} + \gamma \frac{\partial}{\partial x_4'} \right) \frac{B_2'}{\gamma} = 0, \tag{31}$$

$$\frac{\partial \left( E_2' + \frac{\beta}{\gamma} B_3' \right)}{\partial x_1'} - \frac{\partial E_1'}{\partial x_2'} + \left( -\beta \frac{\partial}{\partial x_1'} + \gamma \frac{\partial}{\partial x_4'} \right) \frac{B_3'}{\gamma} = 0.$$

Taking into account equation (16) and after the respective simplification in the above formulas is made, the validity of equation (15) is proved and so the covariance of Maxwell equations relative to transformation (23) from [1] of the vacuum field theory becomes a proven fact.

### 2. Doppler effect

Let us assume that the motion of frame $K'$ relative to frame $K$ is accomplished in the standard way. At a moment of time $t = t' = 0$ when the origin $O$ of frame $K$ coincides with the origin $O'$ of frame $K'$, the latter emanates a plane electromagnetic wave, the front of which is in plane $OXY \equiv O'X'Y'$. Let us assume that point $P$ in which the wave is registered is immovable relative to frame $K$ and is situated in plane $OXY$.

The phases of the electromagnetic wave along the trajectory $OP$ should be identical both relative to frame $K$ and frame $K'$, i.e.

$$\nu \left( \frac{x \cos \vartheta + y \sin \vartheta}{c} - t \right) = \nu' \left( \frac{x' \cos \vartheta' + y' \sin \vartheta'}{c_0'} - t' \right). \tag{32}$$

After substituting $x'$, $y'$, $t'$ with the values equal to them, in accordance with transformation (23) from [1] in the above formula, and equalizing the coefficients in front of $x$, $y$, $t$ in it, the following equations are obtained

$$\frac{\nu \cos \vartheta}{c} = \frac{\nu' \cos \vartheta'}{c_0'}, \quad \frac{\nu \sin \vartheta}{c} = \frac{\nu' \sin \vartheta'}{c_0'}, \quad \nu = \gamma \nu' \left( 1 + \frac{\mathrm{v} \cos \vartheta'}{\gamma c_0'} \right). \tag{33}$$

These formulas lead to

$$\vartheta = \vartheta', \quad \frac{\nu'}{\nu} = \frac{c_0'}{c}, \quad c_0' = \frac{c \left( 1 - \frac{\mathrm{v}}{c} \cos \vartheta \right)}{\gamma} \tag{34}$$

and, consequently, the observed frequency $\nu$ in frame $K$ will be equal to

$$\nu = \nu_0 \left( \frac{\sqrt{1 - \frac{\mathrm{v}^2}{c^2}}}{1 - \frac{\mathrm{v}}{c} \cos \vartheta} \right), \tag{35}$$

where $\nu_0 = \nu'$ is the natural frequency of the source in frame $K'$.

The latter formula coincides with Doppler effect formula determined in special relativity and experimentally checked by Ives and Stillwell [2,3] through a simultaneous observation of the emanation from one and the same source in two opposite directions.

### 3. Real aberration

The term real aberration will be used to designate conventionally the deviation of light in the transition from one inertial reference system to another inertial reference system, moving in relation to the first one with a specified velocity. Let us consider the case in which the plane electromagnetic wave is emanated in frame $K'$ and after that it propagates in the proper space of frame $K$ with velocity $c$ until it reaches the point $P$, immovably connected with it.

Let frame $K'$ move in relation to frame $K$ in the standard way and at the moment of time $t = t' = 0$ at which the origin $O$ of frame $K$ coincides with the origin $O'$ of frame $K'$. A plane electromagnetic wave emanates from origin $O'$ with a front in plane $OXY \equiv O'X'Y'$. The real aberration will be determined using the fronts of the waves which in the frames $K$ and $K'$ are equal to

$$d\Phi = \frac{\vec{k} \cdot d\vec{r}}{c} - \omega \, dt, \tag{36}$$

$$d\Phi' = \frac{\vec{k}' \cdot d\vec{r}'}{c'} - \omega' \, dt', \tag{37}$$

where $\omega$ and $\omega'$ are the frequencies of the waves, $\vec{r}$ and $\vec{r}'$ are their radius vectors, and $\vec{k}$ and $\vec{k}'$ are their wave vectors in frames $K$ and $K'$. The latter are equal to

$$\vec{k} = \frac{\omega}{c}\vec{n}, \quad \vec{k}' = \frac{\omega'}{c}\vec{n}', \tag{38}$$

where $\vec{n}$ and $\vec{n}'$ are the single vectors in the direction of wave propagation in frames $K$ and $K'$.

In this case

$$d\vec{r} = \vec{r}(t+dt) - \vec{r}(t), \quad d\vec{r}' = \vec{r}'(t'+dt') - \vec{r}'(t'), \tag{39}$$

which as can be seen in Fig. 1, gives

$$d\vec{r} = dr\vec{n}, \quad d\vec{r}' = dr'\vec{n}', \tag{40}$$

where $dr$ and $dr'$ are the magnitudes of the elementary vectors $d\vec{r}$ and $d\vec{r}'$. Consequently formulas (36) and (37) can be presented in the form

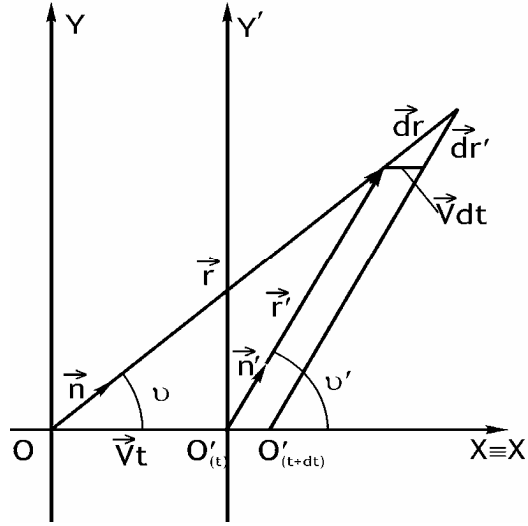

**Fig. 1.**

$$d\Phi = \omega\left(\frac{dr}{c} - dt\right) = 0, \quad d\Phi' = \omega'\left(\frac{dr'}{c'} - dt'\right) = 0. \tag{41}$$

Since the wave frequencies $\omega$ and $\omega'$ differ from zero, the above formulas will be valid in case

$$\frac{dr}{c} - dt = 0, \quad \frac{dr'}{c'} - dt' = 0. \tag{42}$$

The front of the wave emanated from the origin $O'$ of frame $K'$ at the moments of time $t$ and $t+dt$, is presented in Fig. 1. To the elementary period of time $dt$ in frame $K$ there corresponds an elementary period of time $dt'$ in frame $K'$ equal to

$$dt' = \gamma \, dt. \tag{43}$$

As presented in Fig. 1, the following formulas are valid

$$dr \sin \vartheta = dr' \sin \vartheta', \quad dr \cos \vartheta = vdt + dr' \cos \vartheta'. \tag{44}$$

The result of the three formulas above is

$$c \sin \vartheta = c'\gamma \sin \vartheta', \quad c \cos \vartheta = v + c'\gamma \cos \vartheta' \tag{45}$$

out of which follows

$$c' = \frac{c}{\gamma}\sqrt{1 - 2\beta \cos \vartheta + \beta^2}, \quad \sin \vartheta' = \frac{\sin \vartheta}{\sqrt{1 - 2\beta \cos \vartheta + \beta^2}}. \tag{46}$$

The difference in formula for $\sin \vartheta'$ in special relativity and the vacuum field theory equals to $\frac{\beta^2}{2}\sin \vartheta' \cos^2 \vartheta'$. As is known, the star angles of an order of magnitude of $\beta^2$, being too small, cannot be observed experimentally.

Let us determine the amplitude of the plane electromagnetic wave in frame $K$ in case it is emanated by a source, immovable in frame $K'$. Let the amplitude of the above-mentioned wave in the proper space $V_{k'}$ of frame $K'$ be $A'$ and its electric field oscillate in plane $O'X'Y'$, while its magnetic field has a component along axis $O'Z'$. The components of the electric and magnetic field of the wave will be equal to

$$E'_1 = E'_x = E' \sin \vartheta', \quad E'_2 = E'_y = E' \cos \vartheta', \quad E_3 = 0, \tag{47}$$

$$H'_1 = H'_x = 0, \quad H'_2 = H'_y = 0, \quad H'_3 = H'_z = H'. \tag{48}$$

Using formulas (18) – (21) the components of the electric and magnetic field of the wave in frame $K$ will be

$$E_1 = E_1' = E'\sin\vartheta', \quad E_2 = E_2' + \frac{\beta}{\gamma}H_3' = E'\cos\vartheta' + \frac{\beta}{\gamma}H', \tag{49}$$

$$E_3 = E_3' - \frac{\beta}{\gamma}H_2' = 0, \quad H_1 = H_1' = 0, \quad H_2 = H_2' = 0, \quad H_3 = H_3' = H'. \tag{50}$$

As is known, the amplitudes of the electric and the magnetic fields of the wave are equal. In accordance with the covariance of Maxwell equations in the transition from frame $K$ to frame $K'$, to the wave in frame $K$, propagating in its proper space with speed $c$ and amplitude $A$ corresponds a wave in frame $K'$ with an amplitude $A'$. The result from the consideration of the above-mentioned and the usage of the above formulas for the amplitude $A$ of the wave, emanated in frame $K'$ with amplitude $A'$, is

$$A = \sqrt{E_1^2 + E_2^2} = A'\sqrt{1 + 2\frac{\beta}{\gamma}\cos\vartheta' + \frac{\beta^2}{\gamma^2}}. \tag{51}$$

### 4. Observable aberration

The term observable aberration is used to denote the deviation of light due to its transition from the proper space $V_k$ of frame $K$ to the proper space $V_{k'}$ of frame $K'$ or the reverse transition. In the first of the two cases after the light is emitted in frame $K$ at a certain angle, the change in its direction can be registered by an observer, positioned in the proper space $V_{k'}$ of frame $K'$. On those grounds, the aberration can be called observable. Let us consider the first case, in which we assume that $O'X'Y'$ is a plane of symmetry for $V_{k'}$ and plane $O'Y'Z'$ is its boundary plane. The space $V_k$ from the origin $O$ to plane $O'Y'Z'$ will be considered a proper space of frame $K$, in which the speed of light, measured in frame $K$ equals $c$.

Let us assume that the stipulations, accepted for Doppler effect are valid for the motion of frame $K'$ relative to frame $K$. A light beam is emanated towards frame $K'$ from the origin $O$ of frame $K$ at a sharp angle $\vartheta$ to axis $OX$, propagating in plane $OXY$. The plane $OXY$ coincides with plane $O'X'Y'$. The propagation speed of the light beam relative to frame $K$ is $c$, and relative to frame $K'$, it is $c_0'$. The speed $c_0'$ can be calculated using the last of (34) formulas. After passing in the proper space $V_{k'}$ of frame $K'$, the speed of the beam becomes equal to $c$ and, as a result, it is refracted by the boundary plane at an angle $\vartheta'$ that satisfies the formula below:

$$\frac{\sin\vartheta}{\sin\vartheta'} = \frac{c_0'}{c} = \frac{1 - \frac{v}{c}\cos\vartheta}{\sqrt{1 - \frac{v^2}{c^2}}}. \tag{52}$$

This formula for the aberration coincides with the one deduced in special relativity and presented in [4].

### 5. Reflection from a moving mirror

Let us investigate the reflection of a plane electromagnetic wave from a moving mirror when the ideal mirror surface coincides with the boundary of the proper space of frame $K'$ and the falling and reflected wave are in the proper space of frame $K$. The movement of frame $K'$ relative to frame $K$ will be considered accomplished in the standard way given in [1] Sec.1.

The first case to be discussed is the case when the mirror is immovable relative to frame $K'$ and its mirror surface coincides with plane $O'X'Y'$ while the trajectory of the light beam is in plane $O'X'Z'$. In frame $K$, $\vartheta_1$ and $\vartheta_2$ are the angles between axis $OX$ and the light beams, characterizing the falling and the reflected wave, $v_1$, $v_2$, and $A_1$, $A_2$ are their frequencies and amplitudes. In frame $K'$ the corresponding parameters are $\vartheta_1'$, $\vartheta_2'$, $v_1'$, $v_2'$ and $A_1'$, $A_2'$. The final result from the investigation of the case treated, presented in details in [6], is

$$\vartheta_2 = -\vartheta_1. \tag{53}$$

This indicates that the reflection law in the case discussed does not differ from the reflection law for the immovable mirror. This discussion makes it clear that, in the case of reflection only, the velocity component of the mirror, parallel to the normal to its surface, is of importance. Therefore, for the second case let us assume that for the same motion of frame $K'$ the mirror surface coincides with the plane $O'Y'Z'$ of frame $K'$. It is obvious that the mirror velocity $\bar{v}$ will coincide with the inner normal to its surface and its direction - with the positive direction of axis $O'X'$. The final formula of the amplitude $A_2$ of the reflected wave in frame $K$, obtained on the grounds of the investigations, is

$$A_2 = A_1 \sqrt[4]{1 - 2\beta\cos\vartheta_1 + \beta^2} \cdot \sqrt{(1+\beta^2)\sqrt{1 - 2\beta\cos\vartheta_1 + \beta^2} - 2\beta\cos\vartheta_1 + 2\beta^2},$$

$$p = \frac{A_1^2}{8\pi \beta}\{\cos\vartheta_1 - \beta - \sqrt{1-2\beta\cos\vartheta_1+\beta^2}.$$
$$\cdot[(1+\beta^2)\sqrt{1-2\beta\cos\vartheta_1+\beta^2} - 2\beta\cos\vartheta_1+2\beta^2].(\beta + \frac{(1+\beta^2)\cos\vartheta_1-2\beta}{1-2\beta\cos\vartheta_1+\beta^2})\}. \tag{54}$$

For the amplitude $A_{2s}$ of the reflected wave in special relativity in accordance with [5] the following formula is valid

$$A_{2s} = A_1 \frac{1-2\beta\cos\vartheta_1+\beta^2}{1-\beta^2}. \tag{55}$$

The formulas of $\cos\vartheta_2$ and $v_2$, obtained in the vacuum field theory, are

$$\cos\vartheta_2 = -\frac{(1+\beta^2)\cos\vartheta_1-2\beta}{1+\beta^2-2\beta\cos\vartheta_1}, \quad v_2 = v_1\frac{1-2\beta\cos\vartheta_1+\beta^2}{(1-\beta)^2}. \tag{56}$$

The two formulas above coincide with those derived in special relativity [5]. The difference between the amplitudes of the reflected wave in the two theories, if we limit our consideration to an order of magnitude of $\beta^2$, is equal to $\frac{5}{8}A_1\beta^2\cos^2\vartheta_1$.

## 6. Light pressure

Let us consider the light pressure $p$ acting on the surface of the mirror for the latter case from the previous section, using the symbols for the physical quantities and the angles accepted there. The energies of the falling and the reflected light for unit of area $W_1$ and $W_2$ are equal to

$$W_1 = \frac{cA_1^2}{8\pi}(\cos\vartheta_1-\beta), \quad W_2 = \frac{cA_2^2}{8\pi}(-\cos\vartheta_2+\beta). \tag{57}$$

The difference between the $W_1 - W_2$ is equal to $pv$ and thereof, taking into account formulas (54) and (56), the result for $p$ will be

$$p = \frac{A_1^2}{8\pi \beta}\{\cos\vartheta_1 - \beta - \sqrt{1-2\beta\cos\vartheta_1+\beta^2}.$$
$$\cdot[(1+\beta^2)\sqrt{1-2\beta\cos\vartheta_1+\beta^2} - 2\beta\cos\vartheta_1+2\beta^2].(\beta + \frac{(1+\beta^2)\cos\vartheta_1-2\beta}{1-2\beta\cos\vartheta_1+\beta^2})\}. \tag{58}$$

The first approximation for $p$ is equal to the formula well-known from other theories and is experimentally proved

$$p = \frac{A_1^2}{4\pi}\cos^2\vartheta_1, \tag{59}$$

which is well-known from other theories and is experimentally proved.

## 7. Electrodynamics methods used to determine the velocity of an inertial reference system relative to the proper reference system of vacuum

As is known, Doppler effect makes it possible to measure the velocity of the motion of the source of electromagnetic waves and it can be used to determine the velocity of frame $K'$ relative to frame $K$.

Let us present another method that makes it possible to experiment in frame $K'$ itself in order to determine its velocity relative to frame $K$. The ratio $\frac{e}{m_0}$ in frame $K$ satisfies the second of (2.49) formulas from [6]. In frame $K'$ moving with speed v relati to frame $K$, the mass $m'_0$ of the immovable electron in it will be equal to the mass $m$ of the electron moving with speed v relative to frame $K$, i.e.

$$m'_0 = m = \frac{m_0}{\sqrt{1-\frac{v^2}{c^2}}}. \tag{60}$$

Then, in case the electron is moving in the proper space of frame $K'$, using the method of determination of the second of (2.49) formulas from [6] in frame $K$, the following result will be obtained

$$\frac{e}{m'_0} = \frac{e\sqrt{1-\frac{v^2}{c^2}}}{m_0} = \frac{E'}{\rho'H'^2\sqrt{1-\frac{E'^2}{c^2H'^2}}}, \tag{61}$$

where $\rho'$, $E'$ and $H'$ possess the same values in frame $K'$ as the values of $\rho$, $E$, $H$ in frame $K$.

Using this formula for the value of the speed v of the motion of frame $K'$ relative to frame $K$ the following equation is obtained

$$v = c\sqrt{1 - \frac{m_0^2 E'^2}{e^2 \rho'^2 H'^4\left(1-\frac{E'^2}{c^2 H'^2}\right)}}. \tag{62}$$

Besides the problems discussed it is of interest to check the fidelity of the vacuum field theory using the experiments on the theoretically-grounded possibility to reach a speed higher than the speed of light $c$ relative to frame $K$ proposed in [6]. This proved to be possible in the case when light propagates in vacuum of frame $K'$ and the direction of its propagation coincides with the direction of the relative velocity of frame $K'$ relative to frame $K$.

# GRAVITATION IN THE VACUUM FIELD THEORY

## G. Gemedjiev

Plovdiv University Paisii Hilendarski, 24 Tsar Asen Str.4000 Plovdiv, Bulgaria
e-mail: gemedjievg@mail.bg

*The theory of gravitation in the vacuum field theory assumes that the axiom of the activated state of vacuum is valid for the gravitational interaction of material points. It is characteristic for it, that the vacuum state at some distance $r$ from the material point $P'$ is equivalent to that of vacuum, moving relative to it with a certain velocity. This velocity is equal to the velocity a material particle would have at the same distance in case its motion relative to point $P'$ has started at infinity with a zero velocity. A mechanical model of the activated state of vacuum is proposed. This model explains the negative result of Michelson experiment used to determine the velocity of Earth relative to ether. Another basic assumption is that a privileged inertial reference system K exists, relative to which the laws of the gravitational phenomena are valid. The three Newton laws of motion are in force in the vacuum field theory. The other laws valid in it are: the law of equality of inertial and gravitation mass the law of change of material point mass in its motion relative to frame $K$, the law of a material point mass in the gravitation field of another material point and the gravitation law. The motion of a material point, P' the mass of which is $m'$, is investigated in a static spherically symmetric gravitation field. As a result the well-known formula in general relativity, the one concerning the planet perihelion shift and a new phenomena – the change in time of a material point orbit from close to elliptical into circular one are obtained. The integral of energy is investigated in the spherically symmetric gravitation field in a way close to the classical one for the celestial mechanics, as well as momentum and energy of a material point. As a result, the final form of their formulas is obtained. The deflection of light in a static spherically symmetric gravitation field is considered. Assuming that the mass of a photon in its motion in the above-mentioned field remains equal to zero, then for the deflection of light a formula is obtained, identical to that in general relativity. It is shown that the velocity of light during the propagation, under discussion, appears to be smaller than that in a space without gravitation. The result obtained while investigating the delay in the propagation of the electromagnetic wave in a static spherically symmetric gravitation field coincides with the one in general relativity. The final question discussed is the change of light wave frequency in a static spherically symmetric gravitation field and the well-known formula for the process is obtained.*

## 1. Axiom of the activated state of vacuum in a gravitation field

Unlike general relativity the gravitation theory of the vacuum field theory assumes that a privileged inertial reference system exists in which the gravitation phenomena should be treated. This frame, as in the mechanics of the vacuum field theory, is the proper reference system of vacuum $K$, situated in the space, where the material points or the spherically symmetric bodies are situated. Another requirement which frame $K$ must meet is that it must be situated so far away from the interacting objects that their gravitational influence on the time passing in it can be neglected. As for the Solar system, frame $K$ should be situated in a point of its gravitational sphere where the sum of all gravitation and inertial forces is equal to zero.

To explain the nature of the gravitation interaction of the material points, the speed v of a material point $P$ with a mass $m$ in its radial accelerative motion in the gravitational field of the material point $P'$ with mass $m'$ immovable relative to frame $K$, should be determined. Assuming that at an infinite distance the velocity of point $P$ was equal to zero and its mass was $m_0$, and applying the energy conservation law, the speed v of point $P$ in its motion will be

$$v = \sqrt{\frac{2\gamma m'}{r}}, \qquad (1)$$

where $\gamma$ is a gravitational constant and $r$ is the distance between the points $P$ and $P'$.

In [6] supplement 5.2 it has been proved that the field action of locally accelerating systems on a material point $P$, moving in a static spherically symmetric gravitational field, may be substituted with the action of a field of locally inertial frames on the same point. It means that, at any moment of time, we can consider that the gravitational action of the material point $P'$, on the material point $P$, depends both on other physical values and on the velocity of the latter relative to a local inertial frame $K_L$, whose origin coincides with the point $P$ and is moving radially to point $P'$ at a speed v, satisfying the formula (1). In the vacuum field theory we shall also assume that vacuum in frame $K_L$, as in any reference system, is at rest (in an inactivated state) and on moving the frame $K_L$ relative to frame $K$, all the kinematic and dynamic dependences, deduced in the case of frame $K'$, moving relative to frame $K$ in [1] and [2], hold true. Being examined relative to point $P'$, vacuum in frame $K_L$ moves radially relative to it at speed v. Since an inertial frame, possessing the characteristics of frame $K_L$, can be placed at any point of space around point $P'$, we can say that vacuum around point $P'$ is in an activated state. It is equivalent to a radial movement of vacuum to the point $P'$, its speed v satisfying the formula (1) at a distance $r$ from point $P'$.

## 2. Mechanical model of the activated state of vacuum and Michelson experiment

The vacuum field theory proposes a mechanical model of the activated state of vacuum, assuming, that vacuum represents a gas of material $m_v$ particles with dimensions and weight much smaller than those of the atom. It is considered that under gravitation action the $m_v$ particles move radially towards the material point $P'$ with speed v, satisfying formula (1).

The model accepted for the activated state of vacuum in the gravitation field explains to the full the negative result of Michelson experiment [3]. In an equipotential plane, in which the experiment takes place, vacuum is in an uniform activated state, i. e. the $m_v$ particles reach the equipotential plane with uniform velocities parallel to its normal. This proves the impossibility of measuring the velocity of Earth relative to vacuum in that way.

## 3. Gravitation theory laws in the vacuum field theory

The three Newton laws of motion remain in force in the gravitation theory of the vacuum field theory. In their application it should always be borne in mind that the masses of the material points $P$ and $P'$ obey the three laws, mentioned below:

4. The law of equivalency of the inertial and gravitational mass.
5. The law concerning the change of the material point mass in its motion relative to the proper reference system $K$ of vacuum in the space where it is situated. It states: in case a material point $P$ moves relative to frame $K$ at random speed v $\langle$ c its mass $m$ should be equal to

$$m = \frac{m_0}{\sqrt{1 - \frac{v^2}{c^2}}}, \qquad (2)$$

where $m_0$ is the mass of point $P$ when it is at rest relative to frame $K$.

It follows from that law that if at a certain moment of time $t$ the local inertial frame $K_L$ of the material point $P$ moves relative to frame $K$ with a speed $v_L$, then the mass $m'_L$ of the material point $P'$, immovable relative to frame $K_L$, should be

$$m'_L = \frac{m'_0}{\sqrt{1-\frac{v_L^2}{c^2}}}, \qquad (3)$$

where $m'_0$ is the mass of the material point $P'$ when it is immovable relative to frame $K$.

6. The law of the material point mass in the gravitation field of another material point. It states: the mass of the material point $P'$, moving in the gravitational field of the material point $P$ while the distance between them is $r = r(t)$, is equal to

$$m' = \frac{m'_L}{\sqrt{1-\frac{V^2}{c^2}}}, \qquad (4)$$

where $V$ and $m'$ are its speed and mass in the proper space of the local inertial frame $K_L$, the origin of which coincides with point $P'$ and its motion relative to point $P$ is radial and its speed equals

$$v = \sqrt{\frac{2\mu}{r}}, \quad (\mu = \gamma\, m) \qquad (5)$$

where $m$ is the mass of point $P$ in the presence of the gravitational field of point $P'$ while the mass $m'_L$ is calculated using formula (3).

The speed $V$ can be obtained using the results obtained in kinematics of the vacuum field theory. The derivatives of time will be denoted by a point for short.

Let the coordinates of points $P$ and $P'$ and the components of the velocities $\vec{u}$ and $\vec{u}'$ relative to the coordinate system $O_0 XYZ$ of the proper reference system $K$ of vacuum in the space they are situated in, be correspondingly $x$, $y$, $z$, $\dot{x}$, $\dot{y}$, $\dot{z}$ and $x'$, $y'$, $z'$, $\dot{x}'$, $\dot{y}'$, $\dot{z}'$. Then

$$r = \sqrt{(x-x')^2 + (y-y')^2 + (z-z')^2}, \qquad (6)$$

$$\vec{v} = \frac{v}{r}\left[(x-x')\vec{i} + (y-y')\vec{j} + (z-z')\vec{k}\right], \qquad (7)$$

and the speeds $u$ and $u'$ of the points $P$ and $P'$ relative to frame $K$ are

$$u = \sqrt{\dot{x}^2 + \dot{y}^2 + \dot{z}^2}, \qquad (8)$$

$$u' = \sqrt{\dot{x}'^2 + \dot{y}'^2 + \dot{z}'^2}. \qquad (9)$$

Let us connect the coordinate systems $O'_L X'_L Y'_L Z'_L$, $OXYZ$ and $O'X'Y'Z'$ with the proper space of the local inertial frame $K'_L$ and the material points $P$ and $P'$. Their axes are parallel and unidirectional to the corresponding axes of the coordinate system $O_0 XYZ$ of frame $K$. The origins $O$ and $O'$ coincide with points $P$ and $P'$.

In accordance with the last of (23) formulas of [1] the speed $u_0$ of frame $K$ relative to point $P$ will be

$$u_0 = \frac{u}{\sqrt{1-\frac{u^2}{c^2}}}. \qquad (10)$$

The velocity magnitude $v$ of frame $K_L$ is determined relative to point $P$ as well. This makes it possible, using (24), (25) of [1], to obtain the velocity components $\vec{v}_L$ of frame $K_L$ in frame $K$

$$v_{Ln} = v_n \sqrt{1-\frac{u^2}{c^2}} + \dot{n}, \quad (n = x,\ y,\ z) \qquad (11)$$

where

$$v_n = (n - n')\frac{v}{r}. \qquad (12)$$

Using the law of velocity summation (24) of [1], the components of velocity $\vec{V}$ of point $P'$ in frame $K_L$ are obtained

$$V_n = \frac{\dot{n}' - \dot{n} - v_n\sqrt{1-\frac{u^2}{c^2}}}{\sqrt{1-\frac{v_L^2}{c^2}}}, \quad (n = x,\ y,\ z) \qquad (13)$$

where

$$v_L = \sqrt{v_{Lx}^2 + v_{Ly}^2 + v_{Lz}^2}. \tag{14}$$

The formulas (11) and (13) make it possible, in accordance with law 5 and this law, to obtain for the mass $m'$ of point $P'$ in the gravitational field of point $P$ the following expression

$$m' = \frac{m'_L}{\sqrt{1-\frac{V^2}{c^2}}} = \frac{m'_0}{\sqrt{\left(1-\frac{V^2}{c^2}\right)\sqrt{1-\frac{v_L^2}{c^2}}}}, \tag{15}$$

where

$$V = \sqrt{V_x^2 + V_y^2 + V_z^2}. \tag{16}$$

Due to the total symmetry of the gravitational interaction of two material points, it is obvious, that formula (15) will be valid for the mass $m$ of the material point $P$ in the gravitational field of the material point $P'$ and formulas (11) and (13) specify the magnitudes of its parameters. For that purpose, in all of them the primed quantities exchange places with the non-primed quantities. In accordance with this the mass $m$ will be

$$m = \frac{m_L}{\sqrt{1-\frac{V'^2}{c^2}}} = \frac{m_0}{\sqrt{\left(1-\frac{V'^2}{c^2}\right)\sqrt{1-\frac{v_L'^2}{c^2}}}}. \tag{17}$$

In scientific investigations it is enough to determine the mass with an accuracy with an order of magnitude $\frac{v^2}{c^2}$. With the above-mentioned accuracy, as can be seen from (10), (11), (13) and (15) formulas, the requirement to determine the mass $m$ of particle $P$ in the presence of the gravitational field of point $P'$ drops out.

The last law of the theory of gravitation of the vacuum field theory is the one mentioned, below:

7. Gravitation law. It states: two material points attract each other with a force, the magnitude of which is

$$F = \gamma \frac{m\,m'}{r^2}, \tag{18}$$

where $\gamma$ is the gravitation constant, $m$ and $m'$ are the masses of the points for which formulas (15) and (17) are in force, and $r$ is the distance between them.

### 4. Material point motion in a static spherically symmetrical gravitation field

Let us consider the motion of a material point $P'$ with a mass $m'$ in static spherically symmetrical gravitation field. This motion is equivalent to the its motion in the gravitation field of the material point $P$ with mass $m$, immovable relative to frame $K$.

Let the inertial reference system $K_P$, connected to the material point $P$, consist of a coordinate system $OXYZ$. The origin of the system coincides with point $P$. In the time-measuring system of frame $K_P$ the time interval between the two phenomena is identical to the interval between the same phenomena, measured in frame $K$. The velocity vector of point $P'$ is assumed to lie in plane $OXY$ at a time moment $t = 0$. The equations of motion of the material point $P'$ relative to frame $K_P$ are as follows

$$\frac{d}{dt}\left(m'\frac{dx}{dt}\right) = -\frac{\mu\,m'\,x}{r^3},\quad (\mu = \gamma\,m) \tag{19}$$

$$\frac{d}{dt}\left(m'\frac{dy}{dt}\right) = -\frac{\mu\,m'\,y}{r^3}. \tag{20}$$

The "area velocity" formula at changing mass can be derived from the latter two equations

$$m'\left(x\frac{dy}{dt} - y\frac{dx}{dt}\right) = C_2, \tag{21}$$

where $C_2$ is a constant. On introducing the polar coordinates in the above formula it obtains the form

$$m'r^2\frac{d\theta}{dt} = C_2 = m'_0 r_0^2 \frac{d\theta_0}{dt} = m'_0 \Omega,\quad \left(\Omega = r_0^2 \frac{d\theta_0}{dt}\right), \tag{22}$$

in which the constant $C_2$ is obviously equal to the "area velocity" at a certain moment of time $t_0$, in which the magnitudes of the parameters, present in the above formula, are $m'_0$, $r_0$, $\theta_0$.

It is easy to obtain the formula presented below from equations (19) and (20) by introducing the polar coordinates

$$(m'\dot{r})\dot{} - m'r\dot{\theta}^2 = -\frac{\mu\, m'}{r^2}. \tag{23}$$

In order to obtain the trajectory equation it is necessary to present $r$ as a function of only the angle $\theta$. Excluding $t$ from equation (22) and the two left terms of equation (23), and putting $U = \frac{1}{r}$, the following equation is obtained

$$\frac{d^2 U}{d\theta^2} + U = \frac{m'^2}{m_0'^2} N, \quad \left(N = \frac{\mu}{\Omega^2}\right). \tag{24}$$

In accordance with Sec. 2, in the case discussed, we have

$$V_L = V = \sqrt{\frac{2\mu}{r}}, \quad V_x = \frac{\dot{x} - \frac{vx}{r}}{\sqrt{1 - \frac{v'^2}{c^2}}}, \quad V_y = \frac{\dot{y} - \frac{vy}{r}}{\sqrt{1 - \frac{v'^2}{c^2}}}. \tag{25}$$

Equation (24) can be solved reducing it to an equation of the second order, depending on a small parameter. For the purpose we put

$$U = \frac{1 + ew}{p}, \tag{26}$$

and after the consequent transformation the following equation is deduced

$$\frac{e}{p}\frac{d^2 w}{d\theta^2} + \frac{1 + ew}{p} = \frac{\mu\left(1 - \frac{v_0^2}{c^2}\right)\left(1 - \frac{V_0^2}{c^2}\right)}{\Omega^2\left(1 - \frac{v^2}{c^2}\right)\left(1 - \frac{V^2}{c^2}\right)}, \tag{27}$$

in which $V_0$ and $v_0$ are the magnitudes of $V$ and $v$ at $\theta = \theta_0$ ($\theta_0$ being an integration constant). Putting the equation $\varepsilon = \frac{\mu}{c^2 p}$, and using formulas (25), the following equation is obtained with an accuracy of the order of magnitude $\frac{v^2}{c^2}$

$$\frac{d^2 w}{d\theta^2} + w + \frac{1}{e} = \frac{\mu p}{e\Omega^2} + \varepsilon\left[6w - 2\sqrt{2}\sqrt{1+ew}\frac{dw}{d\theta} + ew^2 + \right.$$
$$\left. + e\left(\frac{dw}{d\theta}\right)^2 - 6w_\alpha + 2\sqrt{2}\sqrt{1+ew}_\alpha \frac{dw_\alpha}{d\theta} - ew_\alpha^2 - e\left(\frac{dw_\alpha}{d\theta}\right)^2\right]. \tag{28}$$

The motion of the material point $P'$ should have been elliptical if the activated state of vacuum in the gravitation field had not influenced its motion. On this basis the solution of equation (28) should be sought in the form

$$w = b(\theta)\cos\psi(\theta) + \varepsilon d(\theta), \tag{29}$$

where the first approximations of the functions $b(\theta)$ and $\psi(\theta)$ relative to the small parameter $\varepsilon$ should be presented in the form

$$b(\theta) = 1 + \varepsilon f(\theta), \tag{30}$$
$$\psi(\theta) = \omega_1(\theta - \theta_0) + \alpha, \quad (\omega_1 = 1 + \sigma\varepsilon), \tag{31}$$

in which $\alpha$ and $\sigma$ are constants.

Replacing these approximations of the functions $b(\theta)$ and $\psi(\theta)$ in equation (28) and equating the terms in front of the zero degree $\varepsilon$, the following is obtained for $\Omega$

$$\Omega = \sqrt{\mu p}, \tag{32}$$

and this equation becomes a differential equation with a small parameter $\varepsilon$ in the form

$$\frac{d^2 w}{d\theta^2} + \omega^2 w = \varepsilon f\left(w, \frac{dw}{d\theta}\right). \tag{33}$$

In accordance with the approximate method of Krilov and Bogoliubov, used to solve it and presented in [4], its first approximation is presented in the form

$$w = b\cos\psi, \tag{34}$$

and it is accepted that

$$\frac{dw}{d\theta} = -b\omega \sin\psi. \qquad (35)$$

After equation (33) is solved using the above-mentioned method, the parameters $b$ and $\psi$ will be presented by the following expressions

$$b = \frac{b_0}{\sqrt{\left(1 - \frac{b_0^2 e^2}{32}\right) ehp\left(2\sqrt{2}\varepsilon\,(\theta-\theta_0)\right) + \frac{b_0^2 e^2}{32}}}, \quad \psi = (1-3\varepsilon)(\theta-\theta_0) + \alpha, \qquad (36)$$

where $b_0$, $\theta_0$ and $\alpha$ are integration constants. If it is assumed, that when $\theta = \theta_0$, the eccentricity would have been equal to $e$, we must put $b_0 = 1$. Based on physical considerations, the angle $\alpha$ is equal to minus the length of the perihelia $\pi$.

The approximate method permits the usage of a specification that will guarantee that all terms linear relative to $\varepsilon$ will be taken into consideration. After accomplishing the above-mentioned specification, the following formula for $r$ is obtained

$$r = \frac{a(1-e^2)}{1 + \sqrt{2}\,b\,e\varepsilon\left(\frac{be}{6} + \frac{b^2 e^2}{128}\right) - \varepsilon e\left(b^2 + g(\theta_0)\right) + be\cos\psi}. \qquad (37)$$

It is obvious from the first of (36) formulas, that when $\theta$ tends to infinity, $b$ tends to zero. Taking this into account, it becomes obvious from (37), that when the angle $\theta$ tends to infinity, the trajectory of the material point $P'$ tends to a circle with a radius $r$ equal to $\dfrac{a(1-e^2)}{1-\varepsilon\,e\,g(\theta_0)}$.

Formula (37) can obtain the following normal form

$$r = \frac{a'(1-e'^2)}{1 + e'\cos\psi}, \qquad (38)$$

where

$$e' = \frac{be}{1 + \sqrt{2}\,b\,e\varepsilon\left(\frac{be}{6} + \frac{b^2 e^2}{128}\right) - \varepsilon e\left(b^2 + g(\theta_0)\right)}, \qquad (39)$$

$$a' = \frac{a(1-e^2)}{\left(1 + \sqrt{2}\,b\,e\varepsilon\left(\frac{be}{6} + \frac{b^2 e^2}{128}\right) - \varepsilon e\left(b^2 + g(\theta_0)\right)\right)(1-e'^2)}. \qquad (40)$$

The numerical calculation of the change in eccentricity of a material point $P'$ for a century when the parameters of the orbit are identical with those of Earth and the angle $\pi = 0$, presuming that the motion started on 1st January 1900, and using (37), gives

$$e' - e = -1{,}4704 \cdot 10^{-7}. \qquad (41)$$

This change is considerably smaller than the change of the eccentricity of Earth for the same period when in its motion relative to the Sun it is treated as a material body of finite mass and in accordance to [5] it amounts to $-4{,}1926 \cdot 10^{-5}$.

## 5. Energy integral

By multiplying (19) and (20) equations, respectively, by $m'\dot{x}$ and $m'\dot{y}$ and then summing them, we obtain

$$\frac{1}{2}\left(m'^2(\dot{x}^2 + \dot{y}^2)\right)^{\cdot} = -\frac{m'^2 \mu (x^2 + y^2)^{\cdot}}{2r^3} = m'^2 \mu \left(\frac{1}{r}\right)^{\cdot}. \qquad (42)$$

The above formula in polar coordinates writes

$$\frac{\dot{m}'}{m'}(\dot{r}^2 + r^2\dot{\theta}^2) + \frac{1}{2}\left(\dot{r}^2 + r^2\dot{\theta}^2\right)^{\cdot} = \mu\left(\frac{1}{r}\right)^{\cdot}. \qquad (43)$$

After some transformations and integration, the energy integral with an accuracy of the order of magnitude of $\dfrac{v^2}{c^2}$ will be

$$\frac{1}{2}(\dot{r}^2 + r^2\dot{\theta}^2) = \mu\left(\frac{1}{r}\right) - \frac{\mu}{2c^2}(v^2 + u^2)\left(\frac{2}{r} - \frac{1}{a}\right) - \frac{\mu}{c^2}\int \frac{(v^2 + u^2)}{r^2}\frac{dr}{d\psi}\frac{d\psi}{dt}dt + C, \qquad (44)$$

where $C$ is an integration constant.

## 6. Momentum and energy of a material point in a static spherically symmetric gravitational field

In finding momentum and energy of a material point $P$ in a static spherically symmetric gravitational field we shall make use of the spatial arrangement of the two frames $K$, $K_L$ and the central body $S$, given in Fig. 1. In this figure the velocity $\vec{v}$ of point $P$ lies in the plane $O_L X_L Y_L$, coinciding with the plane $OXY$. The components of the velocity $\vec{V}$ relative to frame $K$ are:

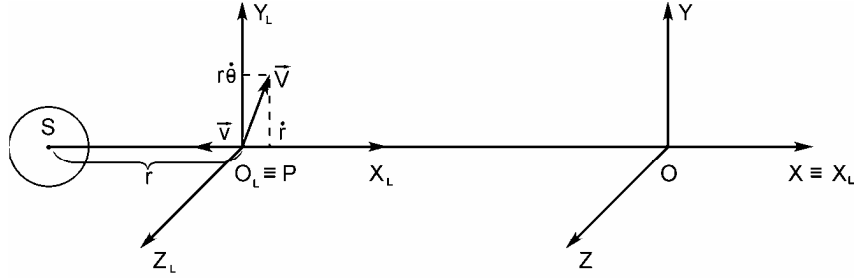

**Fig. 1.**

$$V_x = \dot{r},\ V_y = r\dot{\theta},\ V_z = 0. \tag{45}$$

Using the velocity summation law (24) from [1] for the components of velocity $\vec{V}_L$ and its magnitude $V_L$ relative to frame $K_L$, we find:

$$V_{Lx} = \frac{\dot{r} - \mathrm{v}}{\sqrt{1 - \frac{\mathrm{v}^2}{c^2}}},\ V_{Ly} = \frac{r\dot{\theta}}{\sqrt{1 - \frac{\mathrm{v}^2}{c^2}}},\ V_{Lz} = 0, \tag{46}$$

$$V_L = \sqrt{V_{Lx}^2 + V_{Ly}^2} = \sqrt{\frac{\dot{r}^2 + r^2\dot{\theta}^2 + \mathrm{v}^2 - 2\mathrm{v}\dot{r}}{1 - \frac{\mathrm{v}^2}{c^2}}},\ \left(\mathrm{v} = \sqrt{\frac{2\gamma\, m_s}{r}}\right), \tag{47}$$

where $m_s$ is the mass of the body S.

Making use of the fifth law of gravitation for momentum $p_L$ and energy $E_L$ of point $P$ relative to frame $K_L$, we obtain

$$p_{Lx} = \frac{m_0 V_{Lx}}{\sqrt{\left(1 - \frac{\mathrm{v}^2}{c^2}\right)\left(1 - \frac{V_L^2}{c^2}\right)}},\ p_{Ly} = \frac{m_0 V_{Ly}}{\sqrt{\left(1 - \frac{\mathrm{v}^2}{c^2}\right)\left(1 - \frac{V_L^2}{c^2}\right)}},\ p_{Lz} = 0, \tag{48}$$

$$E_L = \frac{m_0}{\sqrt{\left(1 - \frac{\mathrm{v}^2}{c^2}\right)\left(1 - \frac{V_L^2}{c^2}\right)}}. \tag{49}$$

Finally, using the formulas of transformation of the components of momentum and energy at transition from frame $K_L$ to frame $K$ (5) from [2], we find

$$p_x = p_{Lx} + \frac{\mathrm{v} E_L}{\sqrt{1 - \frac{\mathrm{v}^2}{c^2}}} = \frac{m_0 \dot{r}}{\left(1 - \frac{\mathrm{v}^2}{c^2}\right)\sqrt{1 - \frac{V_L^2}{c^2}}} = \frac{m_0 \dot{r}}{\sqrt{\left(1 - \frac{\mathrm{v}^2}{c^2}\right)\left(1 - \frac{\dot{r}^2 + r^2\dot{\theta}^2 + 2\mathrm{v}^2 - 2\mathrm{v}\dot{r}}{c^2}\right)}}, \tag{50}$$

$$p_y = p_{Ly} = \frac{m_0 r\dot{\theta}}{\left(1 - \frac{\mathrm{v}^2}{c^2}\right)\sqrt{1 - \frac{V_L^2}{c^2}}} = \frac{m_0 r\dot{\theta}}{\sqrt{\left(1 - \frac{\mathrm{v}^2}{c^2}\right)\left(1 - \frac{\dot{r}^2 + r^2\dot{\theta}^2 + 2\mathrm{v}^2 - 2\mathrm{v}\dot{r}}{c^2}\right)}},\ p_z = 0, \tag{51}$$

$$E = \frac{E_L}{\sqrt{1-\frac{v^2}{c^2}}} = \frac{m_0 c^2}{\left(1-\frac{v^2}{c^2}\right)\sqrt{1-\frac{V_L^2}{c^2}}} = \frac{m_0 c^2}{\sqrt{\left(1-\frac{v^2}{c^2}\right)\left(1-\frac{\dot{r}^2 + r^2\dot{\theta}^2 + 2v^2 - 2v\dot{r}}{c^2}\right)}}. \quad (52)$$

## 7. Deviation of light in a static spherically symmetric gravitation field

The trajectory of the photon in a static gravitation field, created by a spherically symmetric body $S$ with mass $m'$, will be treated in this section. Let us assume, that the mass of the photon at its motion in the above-mentioned field remain equal to zero. In this case its trajectory will be determined only by its motion with speed $c$ in the proper space $V_L$ in the local inertial frame $K_L$, determined in Sec. 1. We shall suppose that the location of the frames $K_L$ and $K$ and the connection of the former to the body $S$ remain the same as in Sec. 1.

Let the light beam propagate in plane $O_0 XY$. Using formulas (25) from [1] and assuming, that the axis $O_0 X$ coincides with vector $\vec{r}$ from the polar coordinates, the square of light speed $c^2$ in the coordinate system $K$ when the light passes the same axis will be

$$c^2 = V_{ix}^2 + V_{iy}^2 = \frac{(V_{ax} - v)^2 + V_{ay}^2}{1-\frac{v^2}{c^2}} = \frac{(\dot{r} - v)^2 + r^2\dot{\theta}^2}{1-\frac{v^2}{c^2}}, \quad (53)$$

hence

$$\dot{r}^2 + r^2\dot{\theta}^2 - 2v\dot{r} + 2v^2 - c^2 = 0. \quad (54)$$

Let us investigate the above equation without the third term since the investigations in [6] p. 48 have proved, that the equation obtained in the above-mentioned way

$$\dot{r}^2 + r^2\dot{\theta}^2 + 2v^2 - c^2 = 0. \quad (55)$$

determines the trajectory of the light beam.

In the us examining of the solution of the above equation we shall replace speed v with its equal from formula (1) its and the equation will be transformed in the following way

$$\dot{r}^2 + r^2\dot{\theta}^2 = \dot{\theta}^2\left(\left(\frac{dr}{d\theta}\right)^2 + r^2\right) = 2\mu\left(\frac{c^2}{2\mu} - \frac{2}{r}\right), \quad (\mu = \gamma m'). \quad (56)$$

The form of this equation shows that the motion discussed is hyperbolic and $r$ in it should be presented in the following form

$$r = \frac{p}{e\cos\theta - 1}, \quad (57)$$

and consequently equation (56) can be presented as

$$r^4\dot{\theta}^2\left(\frac{e^2\sin^2\theta + (e\cos\theta - 1)^2}{p}\right) = r^4\dot{\theta}^2\left(\frac{e^2 - 1 - 2(e\cos\theta - 1)}{p}\right) = r^4\dot{\theta}^2\left(\frac{1}{a} - \frac{2}{r}\right) = 2\mu p\left(\frac{c^2}{2\mu} - \frac{2}{r}\right), \quad (58)$$

where $a = \frac{p}{e^2 - 1}$ is the real semiaxis.

Obviously the solution of equation (58) is

$$r^2\dot{\theta} = \sqrt{2\gamma pM}, \quad a = \frac{2\mu}{c^2}. \quad (59)$$

In the treatment of the hyperbolic trajectory, the following formula is valid for the perihelia distance $q$

$$q = a(e-1). \quad (60)$$

For the case, considered, the Sun radius $R$ will be accepted as a perihelion distance. Then the form of the two latter formulas will be

$$R = \frac{2\mu}{c^2}(e-1), \quad (61)$$

and hence

$$e = 1 + \frac{c^2 R}{2\mu} \approx \frac{c^2 R}{2\mu}. \quad (62)$$

The angle of deviation of the light beam from its original direction $\Delta$ is equal to the angle between the two asymptotes, and its value is

$$\Delta = \frac{2}{e} = \frac{4\gamma \, m'}{c^2 R} = 1'',7493, \tag{63}$$

where $m'$ is the mass of the Sun.

This magnitude of $\Delta$ is equal to the magnitude, calculated in general relativity, and coincides well with the one determined by the observations of the total Solar eclipses.

## 8. Light velocity in a static spherically symmetric gravitation field

As presented in the previous section the velocity of light $V$ in a static gravitation field, created by a spherically symmetric body $S$ with a mass $m'$ relative to frame $K$, is equal to

$$V = \sqrt{\dot{r}^2 + r^2\dot{\theta}^2} = c\sqrt{1 - \frac{2v^2}{c^2}} \approx c\left(1 - \frac{2\gamma \, m'}{c^2 r}\right) \langle \, c. \tag{64}$$

This formula indicates, that the light velocity relative to frame $K$ in the presence of a gravitational field is smaller compared to that in a space without gravitation.

## 9. Delay of a radar signal in a static spherically symmetric gravitation field

Another effect, connected with the electromagnetic wave in a static spherically symmetric gravitation field is its propagation delay. To observe it, radio waves are sent from a radar, situated on the Earth, to a reflector situated at another point of the Solar system. The reflector sends these waves back to the Earth. Let us investigate the first approximation of the radar signal delay, assuming, that the Earth and the reflector do not rotate, i.e. they are immovable in the static spherically symmetric gravitation field of the Sun. For the first approximation only the delay due to the change in the wave velocity along its straight-line trajectory from the transmitter on the Earth to the reflector and back, accomplished in the gravitation field, would be taken into consideration.

Let us assume, that the Earth transmitter and the reflector lie in the plane $O_0 XY$ of the coordinate system $O_0 XYZ$, the origin of which coincides with the center of the Sun (Fig. 2). The axis $O_0 X$ is selected in such a way that it is parallel to the straight line connecting the transmitter and the reflector. Their coordinates are correspondingly $(-a_T, b)$, $(a_R, b)$. In this case we have

$$y = b, \ \dot{y} = 0, \ r = \sqrt{x^2 + b^2}, \tag{65}$$

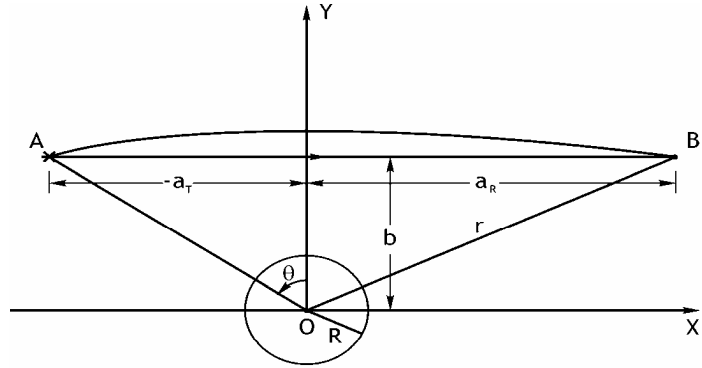

**Fig. 2.**

and as a result of it we obtain and as a result of it we obtain

$$\dot{x}^2 = \dot{r}^2 + r^2\dot{\theta}^2, \quad v^2 = \frac{2\gamma \, m'}{\sqrt{x^2 + b^2}}. \tag{66}$$

Using equation (64) and the latter two formulas, the interval of time $\Delta t$ between the moment the signal is sent and the moment it is received, will be

$$\Delta t = \frac{1}{c} \int_{-a_T}^{a_R} \frac{dx}{\sqrt{1 - \frac{4\gamma \, m'}{c^2 \sqrt{x^2 + b^2}}}}. \tag{67}$$

Due to the inequality $\frac{4\gamma \, m'}{c^2\sqrt{x^2+b^2}} \ll 1$, the above formula with an accuracy of the order of magnitude of $\frac{4\gamma \, m'}{c^2\sqrt{x^2+b^2}}$ can be presented approximately as

$$\Delta t \approx \frac{1}{c}\int_{-a_T}^{a_R}\left(1+\frac{2\gamma \, m'}{c^2\sqrt{x^2+b^2}}\right)dx. \qquad (68)$$

After we do the integration, we obtain

$$\Delta t \approx \frac{1}{c}(a_R+a_T)+\frac{2\gamma \, m'}{c^3}\ln\frac{\left(a_R+\sqrt{a_R^2+b^2}\right)\left(a_T+\sqrt{a_T^2+b^2}\right)}{b^2}. \qquad (69)$$

The interval of time $\Delta\tau$ needed by the signal to go and come back, measured by clocks, situated on the Earth, would be equal to

$$\Delta\tau = 2\Delta t\sqrt{1-\frac{v^2}{c^2}} = 2\Delta t\sqrt{1-\frac{\gamma \, m'}{c^2}\left(\frac{2}{r_z}-\frac{1}{a}\right)}, \qquad (70)$$

where $a$ is the semimajor axis of the Earth orbit, and $r_c$ is the distance between the Sun and the Earth in the period of time $\Delta t$.

Formula (70) coincides with the signal delay formula in general relativity, presented in [7] and in which $c=1$ and $\gamma=1$ are put. Its form is

$$\Delta t \approx \int_{-a_T}^{a_R}\left(1+\frac{(1+\gamma) \, M}{\sqrt{x^2+b^2}}\right)dx, \qquad (71)$$

where $\gamma$ is the "magnitude of space curvature, created by a unit of immovable mass". Using a radar of planets and satellites proved the magnitude of $\gamma$ in general relativity, namely $\gamma=1$, is proved.

From the delay of radio waves in the gravitational field, we can find out whether the space is a Newtonean space or a Riemannean space. A more precise determination of the time-delay in the Solar system can be obtained if it is determined in the system of two satellites, launched in circular orbits around a given planet or around the Sun. The orbits of the two satellites should have different radii and should lie in one and the same plane.

### 10. Light wave frequency change in a static spherically symmetric gravitation field

Let us discuss in frame $K$ the propagation of a light wave of one and the same wave length in a static gravitation field, created by a spherically symmetric body $S$ with a mass $m'$ and a center in point $O_0$ (see Sec. 4) and in a space without a gravitation field. Let us assume, that light propagates along the axis $O_0X$ of frame $K_0$, connected to body $S$. The axis $O_0X$ will be considered coinciding with the axis $OX$ of frame $K$. Due to the fact, that in the vacuum field theory the Newtonean space is accepted, the wave length $\lambda$ of the light wave is uniform relative to frames $K$ and $K_0$.

Let the light wave pass the elementary distance $dx$ in frame $K_0$ for the elementary interval of time $dt_0$, and in frame $K$ for the elementary interval of time $dt$. The magnitudes of the light wave speeds $v$ and $v_0$ in $K$ and $K_0$ frames will be equal to

$$v = \frac{dx}{dt}, \quad v_0 = \frac{dx}{dt_0}. \qquad (72)$$

If the light wave frequencies in $K$ and $K_0$ frames are denoted by $\nu$ and $\nu_o$, then, in accordance with the above, the following formulas will be in force

$$\lambda = \frac{v}{\nu} = \frac{\frac{dx}{dt}}{\nu}, \quad \lambda = \frac{v_0}{\nu_0} = \frac{\frac{dx}{dt_0}}{\nu_0}. \qquad (73)$$

Out of these two formulas and formula (1) with an accuracy of $\frac{\gamma \, m'}{c^2 r}$ the following formula is obtained

$$\nu_0 = \nu\frac{dt}{dt_0} = \frac{\nu}{\sqrt{1-\frac{v^2}{c^2}}} = \frac{\nu}{\sqrt{1-\frac{2\gamma \, m'}{c^2 r}}} \approx \nu\left(1+\frac{\gamma \, m'}{c^2 r}\right), \qquad (74)$$

hence, for the light wave frequency change $\Delta\nu$, the formula, familiar in general relativity for that case, is obtained

$$\Delta\nu = \nu - \nu_0 = -\frac{\gamma \, m'}{c^2 r}\nu. \qquad (75)$$

This formula is checked experimentally with an accuracy of 1% in the gravitation field by Pound and his assistants using the effect of Mössbauer [8,9]